\documentclass{IEEEoj}

\usepackage[T1]{fontenc}
\usepackage{color,array,amsthm}
\usepackage{graphicx}
\usepackage{cite}
\usepackage{amsmath,amssymb,amsfonts}
\usepackage{float}
\usepackage{CJK}
\usepackage{textcomp}
\usepackage{xcolor}
\usepackage{colortbl}
\usepackage{url}
\usepackage{multirow}
\usepackage{graphicx}
\usepackage{hyperref}
\usepackage{subfigure}
\usepackage{wrapfig}
\usepackage{latexsym}
\usepackage{algorithm}
\usepackage{algpseudocode}

\def\BibTeX{{\rm B\kern-.05em{\sc i\kern-.025em b}\kern-.08em
T\kern-.1667em\lower.7ex\hbox{E}\kern-.125emX}}
\AtBeginDocument{\definecolor{ojcolor}{cmyk}{0.93,0.59,0.15,0.02}}
\def\OJlogo{\vspace{-14pt}\includegraphics[height=28pt]{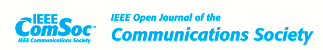}}

\begin{document}
\receiveddate{26 June,2023}
\reviseddate{18 September, 2023}
\accepteddate{8 October, 2023}
\publisheddate{XX Month, 2023}
\currentdate{12 October,2023}
\doiinfo{OJCOMS.2023.1234567}

\title{Latency versus Transmission Power Trade-off in Free-Space Optical (FSO) Satellite Networks with Multiple Inter-Continental Connections}

\author{JINTAO LIANG (Member, IEEE)\textsuperscript{1}}
\author{AIZAZ U. CHAUDHRY (Senior Member, IEEE)\textsuperscript{1}}
\author{JOHN W. CHINNECK\textsuperscript{2}}
\author{HALIM YANIKOMEROGLU (Fellow, IEEE)\textsuperscript{1}} 
\author{GUNES KARABULUT KURT (Senior Member, IEEE)\textsuperscript{3}} 
\author{PENG HU (Senior Member, IEEE)\textsuperscript{4}}
\author{KHALED AHMED\textsuperscript{5}}
\author{STEPHANE MARTEL\textsuperscript{5}}
\affil{Non-Terrestrial Networks (NTN) Lab, Department of Systems and Computer Engineering, Carleton University, Ottawa, ON K1S 5B6, Canada}
\affil{Department of Systems and Computer Engineering, Carleton University, Ottawa, ON K1S 5B6, Canada}
\affil{Poly-Grames Research Center, Department of Electrical Engineering, Polytechnique Montreal, Montreal, QC H3T 1J4, Canada}
\affil{National Research Council of Canada (NRC), Ottawa, ON K1A 0R6, Canada}
\affil{MDA, Sainte-Anne-de-Bellevue QC H9X 3Y5, Canada}

\authornote{This work was supported by the High Throughput and Secure Networks Challenge Program at the National Research Council of Canada.}

\begin{abstract}
In free-space optical satellite networks (FSOSNs), satellites connected via laser inter-satellite links (LISLs), latency is a critical factor, especially for long-distance inter-continental connections.  Since satellites depend on solar panels for power supply, power consumption is also a vital factor. We investigate the minimization of total network latency (i.e., the sum of the network latencies of all inter-continental connections in a time slot) in a realistic model of a FSOSN, the latest version of the Starlink Phase 1 Version 3 constellation. We develop mathematical formulations of the total network latency over different LISL ranges and different satellite transmission power constraints for multiple simultaneous inter-continental connections. We use practical system models for calculating network latency and satellite optical link transmission power, and we formulate the problem as a binary integer linear program. The results reveal that, for satellite transmission power limits set at 0.5 W, 0.3 W, and 0.1 W, the average total network latency for all five inter-continental connections studied in this work levels off at 339 ms, 361 ms, and 542 ms, respectively. Furthermore, the corresponding LISL ranges required to achieve these average total network latency values are 4500 km, 3000 km, and 1731 km, respectively. Different limitations on satellite transmission power exhibit varying effects on average total network latency (over 100 time slots), and they also induce differing changes in the corresponding LISL ranges. In the absence of satellite transmission power constraints, as the LISL range extends from the minimum feasible range of 1575 km to the maximum feasible range of 5016 km, the average total network latency decreases from 589 ms to 311 ms.\\
\end{abstract}



\maketitle

\section{Introduction}
Recent years have seen a rapid development of free-space optical satellite networks (FSOSNs) that are realized using laser inter-satellite links (LISLs) \cite{b1,b2,b3}. Companies such as SpaceX \cite{b4}, Amazon \cite{b5}, and OneWeb \cite{b6} are developing and implementing large low Earth orbit (LEO) satellite constellations, which contain thousands of satellites inter-connected via LISLs \cite{b7}. Since LEO satellite networks have relatively low latency and the capability to provide global coverage \cite{b8,b9,b10}, real-time communication services, like high-frequency trading in financial stock exchanges around the world \cite{b11}, are expected to use FSOSNs to satisfy low-latency requirements. In terms of low latency, high data rate and long-distance communication, the LEO satellite network has major advantages over high orbit satellite networks, such as geostationary Earth orbit satellite networks \cite{b1}, \cite{b12}. According to \cite{b11}, for long-distance communication, when the distance between source ground station and destination ground station is beyond certain value, LEO satellite network shows better latency performance comparing to terrestrial network. In long-distance communication, such as inter-continental connections, multiple satellites are required as relays to complete multi-hop routing from source ground station to destination ground station. Improving the performance of FSOSNs, including latency and power, especially end-to-end latency, is a challenge \cite{b13}, \cite{b14}. 

Latency in a communications network is composed of four kinds of delay: propagation delay, transmission delay, processing delay and queuing delay \cite{b15}. We define node delay to be the summation of processing delay, transmission delay, and queuing delay. The end-to-end latency for an inter-continental connection between a source ground station and a destination ground station through multiple satellites can be calculated by summing up the propagation delay of each link in the connection and the node delay of each satellite in the connection: this is the network latency. We minimize network latency of the FSOSN for multiple simultaneous inter-continental connections via mathematical modelling.

The multi-pair shortest path problem in a network is to compute the shortest paths for multiple specific source-destination pairs simultaneously. In this paper, we propose a binary integer linear program to minimize the total network latency for multiple simultaneous inter-continental connections that span the globe. We build the corresponding mathematical model for total network latency and add realistic constraints based on the characteristics of free-space optical satellite communications. We consider satellite transmission power constraints while minimizing total network latency, since power consumption is a critical factor for LEO satellite networks. We vary the value of the transmission power constraints from 0 to 0.5 W to find out how different satellite transmission power limits affect the total network latency of multiple simultaneous inter-continental connections.

The battery lifetime can directly determine a satellites lifetime, so the longer the battery lasts, the longer the satellite can function. Battery lifetime is determined by its depth of discharge. The less power consumed by a satellite before recharging, the less the depth of discharge of the battery, and the longer the battery lasts. Therefore, reducing or limiting satellite transmission power consumption is important for enhancing a satellites lifetime by extending the life of its battery. For calculating the satellite transmission power for optical uplink/downlink, we consider the losses caused by the atmosphere, including Mie scattering and geometrical scattering \cite{b16}.

In FSOSNs, each satellite has a LISL range to establish reliable connectivity with other nearby satellites. Longer LISL ranges mean that a satellite can connect to more other satellites, and can use these longer connections to e.g. reduce the number of satellites in a route. We study different LISL ranges that vary from the minimum feasible LISL range at 1575 km to the maximum feasible LISL range at 5016 km. We employ the well-known SpaceX Starlink Phase 1 Version 3 constellation \cite{b4}, as it is described in SpaceX’s FCC filings released in 2019. It has 1584 satellites and 22 orbital planes with 72 satellites in each plane. The altitude is 550 km, and the inclination is 53°. Based on the analysis in \cite{b17}, we take the phasing parameter value as 17 for intra-constellation satellite collision avoidance. The Walker constellation notation of this constellation is 53°:1584/22/17. We simulate this constellation based on these parameters. 

We consider five different pairs of inter-continental connections (New York–London, Mexico City–Shanghai, Sao Paulo–Istanbul, Cape Town–Sydney, Cairo–Tokyo), and nine different LISL ranges (1575 km, 1731 km, 2000 km, 2500 km, 3000 km, 3500 km, 4000 km, 4500 km, 5016 km). The results show that for the mathematical formulation to minimize the total network latency (i.e., the sum of the network latencies of all inter-continental connections at a time slot) without the satellite transmission power constraints, longer LISL ranges lead to better total network latency and the minimum total network latency occurs when the LISL range is the maximum feasible value of 5016 km for the Starlink Phase 1 Version 3 constellation. When taking satellite transmission power constraints into consideration, the results show that the average total network latency (i.e., total network latency over 100 time slots) levels off at different LISL ranges for different transmission powers. For example, for a 0.5 W limit on the satellite transmission power, the average total network latency levels off at 339 ms when the LISL range is 4500 km or above. It levels off at lower LISL ranges when the transmission power constraint gets tighter. For a 0.1 W transmission power constraint, the average total network latency levels off at 542 ms when the LISL range is 1731 km or higher. 

The contributions of this work are as follows:
\begin{itemize}
    \item A binary integer linear programming formulation for minimizing total network latency of multiple simultaneous inter-continental connections in FSOSNs under realistic constraints, including satellite transmission power constraints.
    \item An investigation of the total network latency for different simultaneous inter-continental connections at different LISL ranges and different transmission power constraints.
    \item Practical insights on managing FSOSNs, and a list of research challenges for future work.
\end{itemize}

The rest of the paper is organized as follows. The related work and motivation are discussed in Section II. Section III presents the system model, including the network latency model and satellite transmission power models for optical links. Section IV introduces the mathematical formulation to minimize total network latency. Section V discusses the results for network latency of an inter-continental connection at a time slot, total network latency of all inter-continental connections at a time slot, average network latency of an inter-continental connection over 100 time slots, and average total network latency over 100 time slots with and without transmission power constraints, and provides insights based on these results. Conclusions and future work are summarized in Section VI.

\section{Motivation and Related Work}
There are various studies that discuss multi-pair shortest path formulations, satellite network latency minimization, and satellite transmission power as can be seen in \cite{b10,b11,b18,b19,b20,b21,b22,b23,b24,b25,b26,b27,b28,b29,b30,b31,b32,b33,b34,b35} and the references therein.

In \cite{b18}, the multiple pairs shortest path (MPSP) problem in a network is discussed. The authors introduce approaches to compute the shortest paths for specific source-destination pairs. They propose a new algorithm that saves computational work when compared to current solutions for MPSP problems. Their algorithm is effective when shortest paths between specific sets of source-destination pairs must be repeatedly computed using different costs.

In \cite{b19}, the authors introduce several integer programming formulations for the elementary shortest path problem and present a thorough comparison between different formulations. They provide extensive computational experiments and some analytical results, which are very helpful in understanding and implementing an integer programming formulation for the shortest path problem.

In \cite{b20}, the authors formulate and examine the multiple shortest path problem with path deconfliction. They mention several formulations to model multiple source and destination nodes while minimizing both the total distance travelled and path conflict. They present a set of alternative penalty metrics to inhibit path conflict between agents.

In \cite{b21}, the authors discuss the power consumption required when communicating between satellites, and they mention that the battery’s depth of discharge is critical during the communication period. The authors propose several power-control approaches, and their results show that when transmission power is reduced, less depth of discharge is required per cycle, and longer battery lifetime can be achieved.

In \cite{b22}, the authors discuss the power management of satellites in LEO constellations. They mention two critical factors: controlling the satellite transmission power, and the charge and discharge rate of the battery. They propose a satellite discharge model and consider the power consumed for one satellite to transmit packets and the state of charge at the end of the discharge period to improve battery life. This shows that the satellite transmission power can be a crucial factor affecting satellite battery lifetime, and therefore satellite lifetime.

In \cite{b23}, the authors discussed satellite battery life and mentioned that, for long distance missions, the power required per orbit determines the average charging/discharging current for a satellite battery. To remain at long battery lifetime, the average power demanded from a single battery cell should be rather low. In their study, the authors examined the battery of satellites in CubeSats with only body-mounted solar panels, which provide approximately 2 W power generation per 1 U side covered by solar panels. The authors noted that it is important to consider that not all solar panels face the Sun at the same time.

In \cite{b24}, the authors mentioned that for the upcoming satellite networks, power consumption was becoming the major limitation as it affected the battery and thereby the mass and lifetime of the satellite. Since on-board beamforming techniques are severely power-hungry, the authors claimed that the on-board transmission power optimization was becoming a major concern. Therefore, they implemented techniques to reduce the satellite power consumption by reducing the satellite transmission power.

In \cite{b25}, the authors discussed that the power consumption of the satellites during the period when the satellite is in the shadow of the Earth places a heavy load on the satellite's battery and can shorten its lifetime. In addition, the authors mentioned that the life of the battery of an LEO satellite is also influenced by the depth of discharge. The smaller the amount of charge required per cycle, the longer the life of the satellite battery. Therefore, to reduce the depth of discharge (DOD) per cycle, the satellite must control its power consumption.

In \cite{b26}, the authors introduced the mechanism of how satellites charge and discharge in space. They mentioned that when DoD is reduced by 15\%, this doubles the battery life. In order to reduce the power consumption in LEO satellite network, the authors investigated approaches to power down satellite nodes and links during the time when the network traffic is rather low, while still guaranteeing the network connectivity. They explored two heuristic algorithms and evaluated heuristics on a realistic LEO topology and real traffic matrices. 

In \cite{b10}, the authors investigated the network latency of optical wireless satellite networks (OWSNs) versus optical fiber terrestrial networks (OFTNs) for Starlink Phase 1 constellation with an LISL range of 1,500 km. In so doing, they considered three different scenarios for data communications over long-distance inter-continental connections, including connections from New York to Dublin, Sao Paulo to London, and Toronto to Sydney, and they calculated the shortest paths for these inter-continental connections. Their results enabled them to conclude that the OWSN performed better than the OFTN in terms of latency, and the longer the inter-continental distance between source and destination ground stations, the greater the improvement in latency with the OWSN compared to the OFTN.

In \cite{b11}, the authors studied the crossover distance for communicating between two points on Earth beyond which switching or crossing over from an OFTN to an OWSN led to lower latency data communications. They devised a crossover function to calculate the crossover distance and examine the crossover distance in four different scenarios. They simulated three different OWSNs, three different OFTNs, and three different inter-continental connections to study the impact on latency of these networks due to different factors. Their results indicated that the crossover distance varies in accordance with the optical refractive index in OFTNs and the end-to-end propagation distance and satellite altitude in OWSNs.

In \cite{b27}, the authors discuss several latency models for end-to-end networks. They introduce the components of the network path contributing to the total latency, including propagation delay, transmission delay, processing delay, and queuing delay. They also mention an end-to-end latency model that considers inter-satellite links and uplink/downlink.

In \cite{b28}, the authors study the latency minimization of multi-hop satellite links under maximum distance constraints. They propose a nearest neighbor search algorithm to determine the number of hops within the path and the position of the hops. They present numerical results to show that the algorithm has linear complexity, and that the latency performance of their algorithm is close to the minimum latency in an ideal scenario.

In \cite{b29}, the authors simulated Starlink Phase 1 Version 1 constellation and evaluated the satellite network's routing performance with that constellation. They provided an analysis of parameters for this constellation, some insights on the LISL connectivity between satellites in the constellation, and the end-to-end latency properties between two ground stations on Earth. The authors concluded that this satellite constellation has lower latency performance through LISLs than a terrestrial optical fiber network over end-to-end terrestrial distances greater than about 3,000 km.

In \cite{b30}, the authors investigated the low latency communication performance of large satellite constellations, including Starlink and Kuiper. They proposed a new approach for controlling the topology of the satellite network through repetitive patterns, called motifs, and they studied the latency of the satellite network based on this approach. They evaluated the performance of Starlink and Kuiper constellations and showed the improvements on their latency.

In \cite{b31}, the authors discussed queueing and processing delays in a network delay model. Due to the high speed of satellites, inter-satellite links are greatly affected by distance, environment noise, and other factors. To solve the problems of delays and long-distance transmission, they designed a failure probability model for inter-satellite links that considered queuing and processing delays. The authors mentioned the time varying topology for satellite networks and dealt with the dynamic satellite network topology by using a time slice partition method.

In \cite{b32}, the authors discussed several latency models for end-to-end networks. They introduced the components of the network path contributing to the total latency, including propagation delay, transmission delay, processing delay, and queuing delay. They also mentioned the end-to-end latency model that considered inter-satellite link and uplink/downlink.

In \cite{b33}, the authors mentioned the end-to-end delay, propagation delay and network latency in FSOSNs, and studied the effect of LISL range on network latency. They employed the Starlink Phase 1 constellation and studied six different LISL ranges and three different scenarios for long distance communication. They calculated the shortest path between the source and destination ground stations over FSOSN and the network latency for the path. They concluded that longer links lead to better shortest paths with lower network latencies.

In \cite{b34}, the authors discussed the importance of latency in long haul links. They mentioned one major application for low latency communications was high-frequency trading and just a few microseconds of delay could lead to huge loss. They discussed the elements of latency in the links, including the delay for signals to get processed in the electronics and amplifiers, and the delay introduced by optical transmission of the signal through the link. They claimed that the propagation delay is directly proportional to the length of the link, and the longer the link, the greater delay.

In \cite{b35}, the authors introduced the significance of optical inter-satellite links in LEO constellations. They mentioned the inter-satellite links turned the satellite constellation into a mashed network in orbit. They claimed that LISLs were able to address the requirements and provide a size, weight and power advantage compared to traditional radio frequency ISLs.

To the best of our knowledge, for the first time in the literature, we investigate a mathematical formulation to minimize total network latency based on realistic constraints, including satellite transmission power constraints, for the FSOSN arising from Starlink’s Phase 1 Version 3 constellation. This formulation provides accurate results while also being able to handle multiple inter-continental connections simultaneously, and the solution time is rather short.

\section{System Model}
This section introduces the system model for FSOSNs, including network latency models, and satellite transmission power models. 

\subsection{Network Latency Models}
Network latency includes link latency and node latency. As mentioned in \cite{b1}, satellites demand a highly accurate Acquisition, Tracking, and Pointing (ATP) system to establish effective and precise connections with other satellites. This necessity arises from the narrow beam divergence of the laser beam and the varying motion velocities of satellites in space. In the current configuration, the time required for satellites to complete the ATP process, facilitating the establishment of temporary LISLs for node switching, typically requires a few seconds. This delay can be prohibitive for LISLs. Hence, in this study, we have opted to consider node delay while excluding setup delay from our analysis.\\

The link latency is calculated as the propagation delay of an optical link, which depends on the propagation distance of the link, and can be expressed as:
\begin{equation}
\label{eq_1}
T_{link} = d_{prop}/c\text{,}
\end{equation}
where $d_{prop}$ is the propagation distance of the optical link between the transmitter and the receiver, and \textit{c} is the speed of light.

The node latency $\textit{T}_{node}$ for each satellite in the FSOSN is the sum of transmission delay, queuing delay and processing delay. Transmission delay can be very small if the link data rate is very high. For instance, if the link data rate is 10 Gbps and the packet size is 1500 bytes, transmission delay is 1.2 $\mu$s, which is negligible. We assume the data rate as 10 Gbps since this is a practical link data rate for optical satellite communications and the maximum data rate that is offered by Mynaric’s LCT \cite{b45}. Since 1500 bytes is a relatively large packet in satellite communication, we can consider the transmission delay as negligible in this work. According to \cite{b25}, when the router has enough processing capability, it can be assumed that it has approximately equal queuing and processing delays and to simplify the operation, the authors assume queuing delay and processing delay as 4 ms and 6 ms, respectively, in their simulations. We assume node latency to be 10 ms for each satellite in the FSOSN and ignore the transmission delay due to the high link data rate.

\subsection{Satellite Transmission Power Models}
The system models for calculating satellite transmission power in FSOSNs include the optical inter-satellite link transmission power model and the optical uplink/downlink satellite transmission power model.

For an optical inter-satellite link in an FSOSN, the transmission power $\textit{P}_T$ is given by \cite{b36} as
\begin{equation}
\label{eq_2}
P_T = P_R/(\eta_T\eta_RG_TG_RL_TL_RL_{PS}) \text{,}
\end{equation}
where $\textit{P}_T$ is the transmitted power in Watts, $\textit{P}_R$ is the received power in Watts, $\textit{$\eta$}_T$ is the optics efficiency of the transmitter, $\textit{$\eta$}_R$ is the optics efficiency of the receiver, $\textit{G}_T$ is the transmitter gain, $\textit{G}_R$ is the receiver gain, $\textit{L}_T$ is the transmitter pointing loss, $\textit{L}_R$ is the receiver pointing loss, and $\textit{L}_{PS}$ is the free-space path loss for an inter-satellite link. The transmitter gain $\textit{G}_T$ in \eqref{eq_2} is given by \cite{b37} as
\begin{equation}
\label{eq_3}
G_T = 16/(\Theta_T)^{2}\text{,} 
\end{equation}
where $\textit{$\Theta$}_T$ is the full transmitting divergence angle in radians. The receiver gain $\textit{G}_R$ in (2) is given by \cite{b36}
\begin{equation}
\label{eq_4}
G_R = (D_R\pi/\lambda)^2\text{,} 
\end{equation} 
where $\textit{D}_R$ is the receiver’s telescope diameter in mm. The transmitter pointing loss $\textit{L}_T$ in (2) is given as \cite{b36}
\begin{equation}
\label{eq_5}
L_T = \exp{(-G_T(\theta_T)^{2})}\text{,} 
\end{equation}
where $\textit{$\theta$}_T$ is the transmitter pointing error in radians. The receiver pointing loss $\textit{L}_R$ is given as \cite{b36}
\begin{equation}
\label{eq_6}
L_R = \exp{(-G_R(\theta_R)^{2})}\text{,} 
\end{equation}
where $\textit{$\theta$}_R$ is the receiver pointing error in radians. The free-space path loss for inter-satellite link $\textit{L}_{PS}$ is given as \cite{b36}
\begin{equation}
\label{eq_7}
L_{PS} = (\lambda/4\pi\textit{d}_{SS})^2\text{,} 
\end{equation}  
where $\textit{$\lambda$}$ is the operating wavelength in nm, and $\textit{d}_{SS}$ is the propagation distance between the satellites in km.

For optical uplink and downlink in an FSOSN, we consider Mie scattering and geometrical scattering to model atmospheric attenuation. The transmission power $\textit{P}_T$ for optical uplink and downlink is given by \cite{b38}
\begin{equation}
\label{eq_8}
P_T = P_R/(\eta_T\eta_RG_TG_RL_TL_RL_AL_{PG})\text{,}
\end{equation}
where $\textit{L}_A$ is the atmospheric attenuation loss, $\textit{L}_{PG}$ is the free-space path loss for links between ground stations and satellites and the other parameters are as for the optical inter-satellite link transmission power model in (2). 

Mie scattering can be modeled by the following expression \cite{b39}:
\begin{equation}
\label{eq_9}
\rho = a(h_E)^{3}+b(h_E)^{2}+ch_E+d\text{,}
\end{equation}  
where $\rho$ is the extinction ratio, $\textit{h}_E$ is the height of the ground station in km, and $\textit{a}$, $\textit{b}$, $\textit{c}$ and $\textit{d}$ are the wavelength-dependent empirical coefficients, which can be expressed as \cite{b39}
\begin{equation}
\label{eq_10}
\textit{a} = -0.000545\lambda^2+0.002\lambda-0.0038\text{,}
\end{equation}
\begin{equation}
\label{eq_11}   
\textit{b} = 0.00628\lambda^2-0.0232\lambda+0.00439\text{,}
\end{equation}
\begin{equation}
\label{eq_12}   
\textit{c} = -0.028\lambda^2+0.101\lambda-0.18\text{,}
\end{equation}
\begin{equation}
\label{eq_13}   
\textit{d} = -0.228\lambda^3+0.922\lambda^2-1.26\lambda+0.719\text{,}
\end{equation}   
and the atmospheric attenuation due to Mie scattering can be calculated as
\begin{equation}
\label{eq_14}
\textit{I}_m = \exp{(-\rho/\sin({\theta_E}))}\text{,}   
\end{equation} 
where $\textit{$\theta$}_E$ is the elevation angle of the ground station in degrees.

Geometrical scattering can be calculated through the expression below \cite{b39}:
\begin{equation}
\label{eq_15}
\textit{V} = 1.002/(\textit{L}_W\textit{N})^{0.6473}\text{,}   
\end{equation} 
where $\textit{V}$ is the visibility in km, $\textit{L}_W$ is the liquid water content in $g/m^{-3}$ and $\textit{N}$ is the cloud number concentration in $cm^{-3}$. The attenuation coefficient $\textit{$\theta$}_A$ can be expressed as \cite{b39}
\begin{equation}
\label{eq_16}
\theta_A = (3.91/\textit{V}) (\lambda/550)^{-\varphi}\text{,}    
\end{equation}
where $\textit{$\varphi$}$ is the particle size related coefficient given according to Kim’s model. The atmospheric attenuation can be expressed using the Beer-Lambert law as
\begin{equation}
\label{eq_17}
\textit{I}_g= \exp{(-\theta_A\textit{d}_A)}\text{,}    
\end{equation} 
where $\textit{d}_A$ is the distance of the optical beam through the atmosphere and can be expressed as \cite{b40}
\begin{equation}
\label{eq_18}
\textit{d}_A= (\textit{h}_A-\textit{h}_E)\csc({\theta_E})\text{,}    
\end{equation} 
where $\textit{h}_A$ is the height of the troposphere layer of atmosphere in km, $\textit{h}_E$ is the altitude of the ground station in km. 

The atmospheric attenuation loss considering both Mie scattering and geometrical scattering can then be calculated as
\begin{equation}
\label{eq_19}
\textit{L}_A = \textit{I}_m\textit{I}_g = \exp{(-\rho/\sin({\theta_E}))}\exp{(-\theta_A\textit{d}_A)}\text{.}    
\end{equation}

In this work, we add the satellite optical link transmission power as a constraint to the mathematical formulation for minimizing network latency. Thus, we need to calculate the satellite optical link transmission power based on satellite received power, $\textit{P}_R$, which can be calculated based on link margin, $\textit{LM}$, and satellite receiver sensitivity, $\textit{P}_{req}$, through the following equation \cite{b40}:
\begin{equation}
\label{eq_20}
P_R = LM \times P_{req}.
\end{equation}

\section{Mathematical Formulation}
We are interested in minimizing the total network latency of FSOSNs for multiple simultaneous inter-continental connections under realistic constraints, including satellite transmission power constraints. We use a binary integer programming formulation. 

\subsection{Inputs}
For the inputs of the formulation, let
\begin{itemize}
    \item $\textit{S}$ be the set of source ground stations, where $\textit{s}^k$ is the source node for an inter-continental connection $\textit{k}$;
    \item $\textit{D}$ be the set of destination ground stations, where $\textit{d}^k$ is the destination node for an inter-continental connection $\textit{k}$;
    \item $\textit{K}$ be the set of inter-continental connections to be routed, where $\textit{k}$ is the inter-continental connection for source $\textit{s}^k$ and destination ${d}^k$;
    \item ${a}_{ij}$ be the cost of the link ($\textit{i}$, $\textit{j}$) at a LISL range, which is the propagation delay of the link ($\textit{i}$, $\textit{j}$), and it is calculated as in (1). As shown in Figure \ref{figure19}, ${a}_{NY,1}$ refers to the propagation delay of the optical uplink from New York to satellite 1 and ${a}_{1,2}$ refers to the propagation delay of the LISL from satellie 1 to satellite 2; 
\begin{figure*}[htbp]
\centerline{\includegraphics[width=16cm, height=7cm]{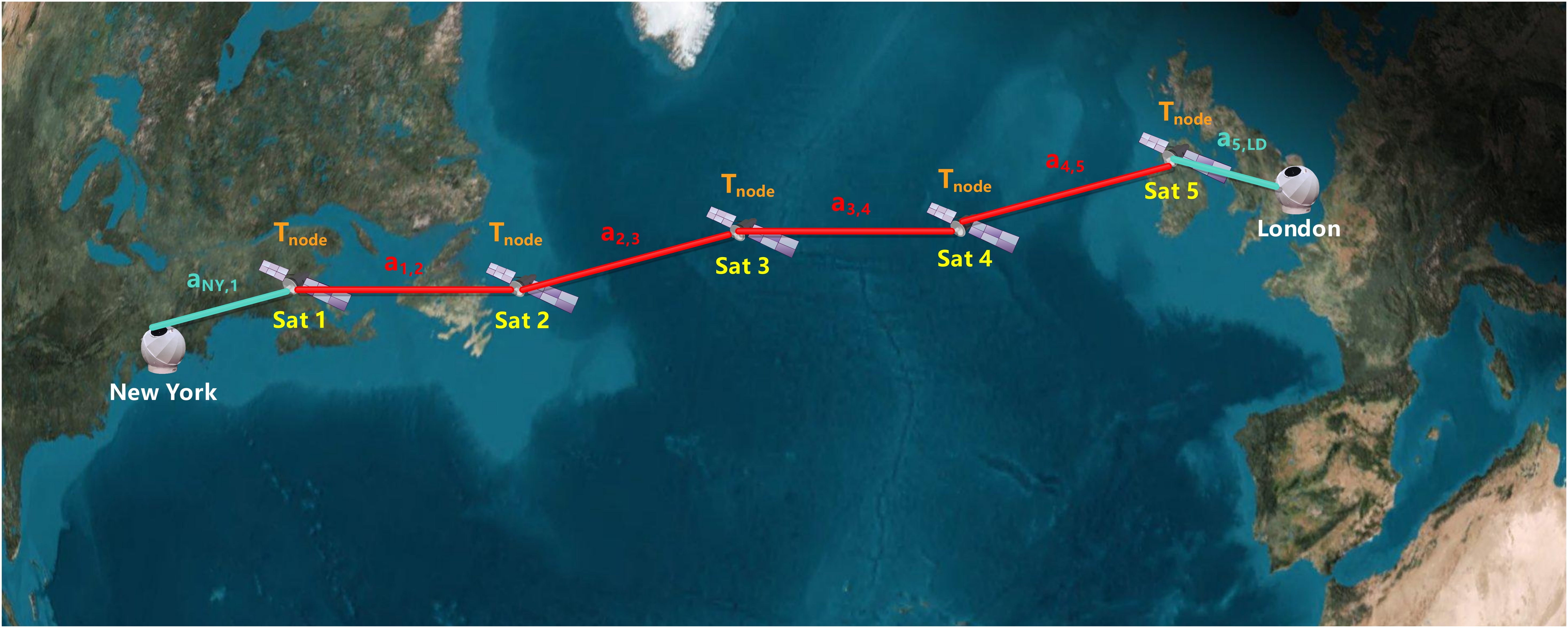}}
\caption{An illustration of the propagation delay of each link and the node delay of each satellite on the path for the inter-continental connection from New York to London.}
\label{figure19}
\end{figure*}
    \item $\textit{b}_{ij}$ be the cost of the link ($\textit{i}$, $\textit{j}$) at a LISL range, which is the satellite transmission power of the optical link ($\textit{i}$, $\textit{j}$) (where this link can be an optical inter-satellite link or an optical uplink/downlink), and is calculated according to the satellite transmission power models in Section III. As shown in Figure \ref{figure20}, ${b}_{NY,1}$ refers to the transmission power of the optical uplink ground station at New York required to transmist data to satellite 1 and ${b}_{1,2}$ refers to the transmission power of the LISL satellite 1 required to transmit data to satellite 2; 
\begin{figure*}[htbp]
\centerline{\includegraphics[width=16cm, height=7cm]{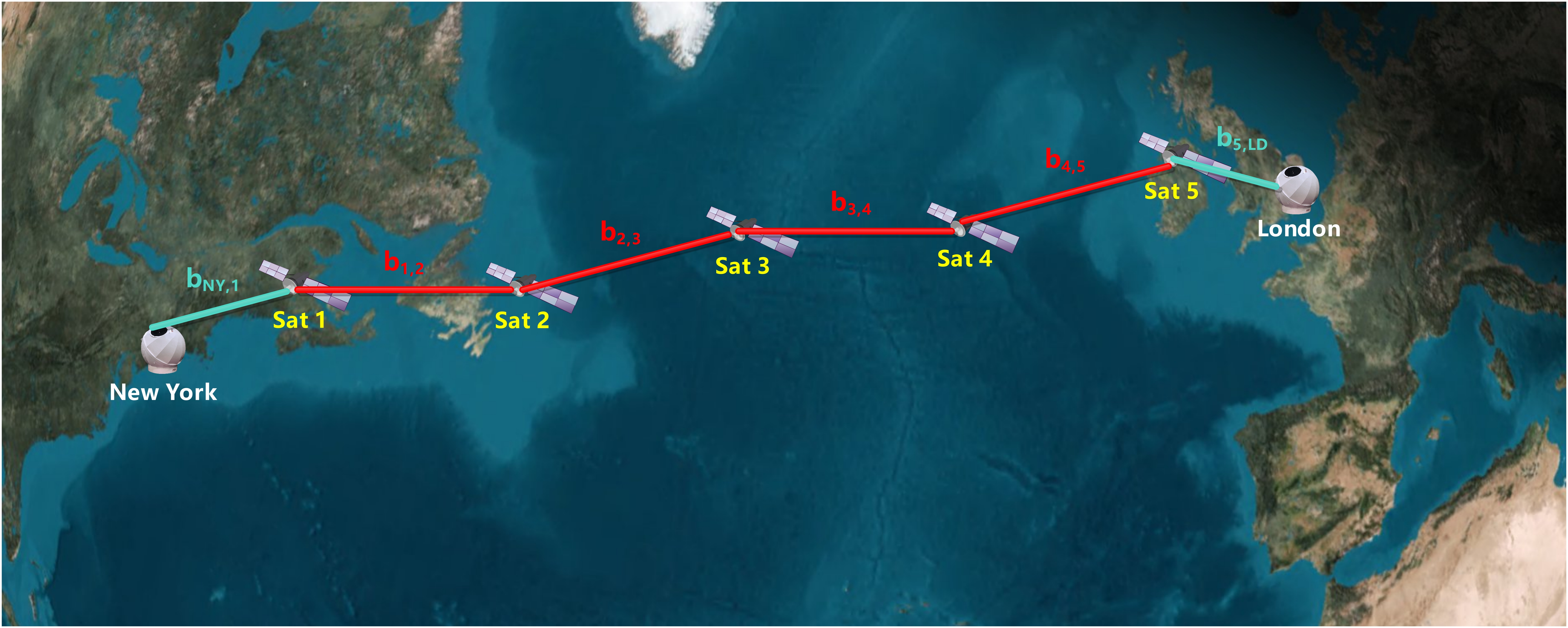}}
\caption{An illustration of the transmission power of each link on the path for the inter-continental connection from New York to London.}
\label{figure20}
\end{figure*}
    \item $\textit{T}_{node}$ be the node latency for each satellite as shown in Figure \ref{figure19}, which is a constant, and is equal to 10 ms;
    \item the node-degree for each satellite be 4; and
    \item ${P}_{lim}$ be the satellite optical link transmission power limit, which is a constant, and is equal to \{0.5 W, 0.3 W, 0.1 W\}. According to \cite{b40}, for an optical inter-satellite link, the transmission power is 0.85 W when the distance between two satellites is 5000 km. Additionally, for an optical uplink/downlink, the transmission power is 0.30 W when the altitude of the satellites is 600 km. Given that we employ Starlink Phase 1 Version 3 constellation in this work, based on the parameters of the constellation we find out that the maximum feasible LISL range is 5016 km and the altitude of the satellites is 550 km. Therefore, we select these three satellite optical link transmission power limits.
\end{itemize}

\subsection{Variables}
For the variables of the formulation, let
\begin{itemize}
    \item $\textit{x}_{ij}^k$ be a binary variable such that $\textit{x}_{ij}^k$ $\in$ \{0,1\} is 1 if the link ($\textit{i}$, $\textit{j}$) belongs to the path for an inter-continental connection $\textit{k}$, and 0 otherwise; and
    \item $\textit{y}_i^k$ be a binary variable such that $\textit{y}_i^k$ $\in$ \{0,1\} is 1 if the node $\textit{i}$ belongs to the path for an inter-continental connection $\textit{k}$, and 0 otherwise.
\end{itemize}

\subsection{Objective}
The objective function can be expressed as\\ 
\begin{equation}
\label{eq_21}
\text{min} \sum_{(i, j)}\sum_{k} \textit{a}_{ij} \textit{x}_{ij}^k + \textit{T}_{node} \sum_{i}\sum_{k} \textit{y}_i^k \text{,}    
\end{equation}                                            
where we minimize the total network latency for all source-destination pairs, i.e., all inter-continental connections, simultaneously. In addition, for each pair (i.e., the source ground station and destination ground station for each inter-continental connection), the solution for (21) provides the network latency (i.e., the summation of the propagation delay of each optical link on the path and the total node delay for all the satellites on the path) from the source ground station to the destination ground station of each inter-continental connection.

\subsection{Constraints}
We consider flow conservation constraints, node-degree constraints, link-disjoint constraints, and satellite optical link transmission power constraints for realistic modeling of FSOSNs.

The following are the flow conservation constraints for the source ground station nodes, the destination ground station nodes, and the intermediate satellite nodes, respectively. Source nodes $\textit{s}^k$ should only send flow; destination nodes $\textit{d}^k$ should only receive flow; and  intermediate satellite nodes both receive and send flow simultaneously:
\begin{equation}
\label{eq_22}
\sum_{j}\textit{x}_{hj}^k - \sum_{i}\textit{x}_{ih}^k = 
\begin{cases} 
    1,       & \quad \text{if} \; h = \textit{s}^k \text{,}\\
    -1,      & \quad \text{if} \; h = \textit{d}^k \text{,}\\
    0,        & \quad \text{otherwise.}
\end{cases}
\end{equation}

The node-degree constraints for all the nodes within the network are as follows. We assume that all satellites have four laser communication terminals \cite{b3}, \cite{b41}, i.e., the total number of incoming links and outgoing links for a satellite node should not be greater than four:
\begin{equation}
\label{eq_23}
\sum_{i} \textit{x}_{ih}^k + \sum_{j}\textit{x}_{hj}^k \leq 4\text{.}\; 
\end{equation}  

The link-disjoint constraints for all the flows within the network are as follows. We assume that a flow over an optical link uses all of the capacity of that optical link, and an optical link between two nodes can only be used by the flow of one inter-continental connection. If a link is being used by one source-destination route, other pairs must find an alternative link to transmit their flows:
\begin{equation}
\label{eq_24}
\sum_{(i,j,k)} \textit{x}_{ij}^k \leq 1\text{.}   
\end{equation}    
The following are the link transmission power constraints for all the optical links in the FSOSN. These restrict the satellite transmission power consumption $P_{lim}$ to one of \{0.5 W, 0.3 W, 0.1 W\}:
\begin{equation}
\label{eq_25}
\textit{b}_{ij}^k\textit{x}_{ij}^k \leq \textit{P}_{lim}\text{, for all}\; \textit{i, j, k.}   
\end{equation}

We do not include constraints on link propagation delay for all the optical links in the FSOSN. The limit is more efficiently implemented by constraining the latency of all links within the specified LISL ranges, and simply omitting optical links with a propagation distance longer than the specified LISL range. 

\section{Numerical Results}

\subsection{Experimental Setup}
We consider five different source ground stations and five different destination ground stations, and five inter-continental connections (New York–London, Cairo–Tokyo,  Sao Paulo–Istanbul, Cape Town–Sydney, and Mexico City–Shanghai). These five connections span all continents and have different end-to-end distances, which makes the results more realistic. Note that the terrestrial distance between source and destination ground stations, i.e., the distance between these ground stations along the surface of the Earth, for the New York–London, Cairo–Tokyo, Sao Paulo–Istanbul, Cape Town–Sydney, and Mexico City–Shanghai inter-continental connections is 5588 km, 9581 km, 10586 km, 11033 km, and 12922 km, respectively. We use Starlink Phase 1 Version 3 constellation for this investigation, which is simulated using the well-known satellite constellation simulator Systems Tool Kit (STK) version 12.1 \cite{b42}. We build the satellite constellation in the STK simulator, add ground stations, and establish optical links between each satellite and ground stations for a given LISL range. We extract this data from STK in Python \cite{b43}, and use it as input to the mathematical model. We solve the binary integer linear program in (21)--(25) using IBM’s commercial solver CPLEX Version 20.1 \cite{b44}.

LISL range is a critical factor in total network latency in FSOSNs. It affects the connectivity of satellites, since a satellite can only establish LISLs with other satellites that are within this range. The maximum LISL range for satellites in a constellation is constrained only by visibility. For the Starlink Phase 1 Version 3 constellation, the satellite altitude is 550 km, and the maximum feasible LISL range can be calculated as 5016 km \cite{b2}. We consider several LISL ranges for satellites (1575 km, 1731 km, 2000 km, 2500 km, 3000 km, 3500 km, 4000 km, 4500 km, and 5016 km). The 1575 km LISL range is a reasonable minimum feasible LISL range for a satellite in this constellation as a satellite can establish six permanent LISLs with other satellites at this range, including four with neighbors in the same orbital plane and two with the nearest left and right neighbors in adjacent orbital planes. Shorter LISL ranges are not sufficient for a satellite to establish the two permanent LISLs with neighbors in adjacent orbital planes. At 1731 km LISL range, a satellite can establish ten permanent LISLs with other satellites, including four with neighbors in the same orbital plane and six with nearest neighbors in adjacent orbital planes. Other ranges, including 2000 km, 2500 km, 3000 km, 3500 km, 4000 km, and 4500 km, are also considered to comprehensively study the effect of LISL range on the total network latency. We obtain the total network latency, which is the sum of the network latencies of the paths for all inter-continental connections, for each LISL range at each time slot. We also obtain the results for 100 time slots and then take the average value of these results over these 100 time slots.

To calculate the satellite optical link transmission power in the FSOSN for the Starlink Phase 1 Version 3 constellation, we use the parameters summarized in Table \ref{table1}, which are used in existing practical optical satellite communication systems. We set the $LM$ for an inter-satellite link as 3 dB and the $LM$ for uplink/downlink as 6 dB since there is more turbulence and attenuation in an optical uplink/downlink. We set the received power ${P}_R$ as -32.5 dBm for an inter-satellite link and -29.5 dBm for an uplink/downlink according to (20). 

\begin{table}
\centering
\caption{Simulation Parameters.}
\label{table1}
\setlength{\tabcolsep}{0.7pt}
\arrayrulecolor{black}
\begin{tabular}{llll} 
\hline
\textbf{Parameter}                          & \textbf{Symbol} & \textbf{Units}          & \textbf{Value}                                                       \\ 
\hline
Laser wavelength\cite{b45}                       & $\textit{$\lambda$}$      & nm                      & 1550                                                                 \\
Transmitter optical efficiency \cite{b46}         & $\textit{$\eta$}_T$     & ~                       & 0.8                                                                  \\
Receiver optical efficiency \cite{b46}            & $\textit{$\eta$}_R$     & ~                       & 0.8                                                                  \\
Data rate \cite{b47}                              & $\textit{R}_{data}$  & Gbps                    & 10                                                                   \\
Receiver telescope diameter \cite{b45}            & $\textit{D}_R$     & mm                      & 80                                                                   \\
Transmitter pointing error \cite{b46}             & $\textit{$\theta$}_T$     & $\mu$rad                    & 1                                                                    \\
Receiver pointing error \cite{b46}               & $\textit{$\theta$}_R$     & $\mu$rad                    & 1                                                                    \\
\begin{tabular}[c]{@{}l@{}}Full transmitting divergence \\ angle \cite{b47}\end{tabular}      & $\textit{$\Theta$}_T$     & $\mu$rad                    & 1.5                                                                  \\
Receiver sensitivity \cite{b47}                   &$ \textit{P}_{req}$   & dBm                     & -35.5                                                                \\
Bit error rate \cite{b47}                         & ~               & ~                       & 10\textsuperscript{-12}   \\
\begin{tabular}[c]{@{}l@{}}Link Margin for \\ inter-satellite link \cite{b40}\end{tabular}      & $ \textit{LM}_{ISL}$     & dB   & 3 \\
\begin{tabular}[c]{@{}l@{}}Link Margin for \\ uplink/downlink \cite{b40}\end{tabular}      & $ \textit{LM}_{UpDown}$     & dB & 6        \\
Ground station attitude                     & $\textit{h}_E$    & km                      & 0.1                                                                  \\
Thin cirrus cloud concentration \cite{b16}         & $\textit{L}_{W}$     & cm\textsuperscript{-3}  & 0.5                                                                  \\
Liquid water content \cite{b16}                    & \textit{N}      & g/m\textsuperscript{-3} & 3.128×10\textsuperscript{-4}                                         \\
Partial size coefficient \cite{b48}               & $\textit{$\Phi$}$      & ~                       & 1.6                                                                  \\
Troposphere layer height \cite{b49}                     & $\textit{h}_A$     & km                      & 20                     \\
\begin{tabular}[c]{@{}l@{}@{}@{}}Location of source \\ ground station\end{tabular}           & \textit{~}      & ~                       & \begin{tabular}[c]{@{}l@{}@{}@{}@{}}\{New York, \\Cairo, \\Sao Paulo, \\Cape Town, \\ Mexico City\} \end{tabular}  \\
\begin{tabular}[c]{@{}l@{}@{}@{}}Location of destination \\ ground station\end{tabular}           & \textit{~}      & ~                       & \begin{tabular}[c]{@{}l@{}@{}@{}@{}}\{London, \\Tokyo, \\Istanbul, \\Sydney, \\ Shanghai\} \end{tabular}  \\
\begin{tabular}[c]{@{}l@{}@{}@{}}LISL range for Starlink Phase 1 \\ Version 3 \end{tabular}& \textit{~}      & km                      & \begin{tabular}[c]{@{}l@{}@{}@{}@{}}\{1575, 1731, \\ 2000, 2500, \\3000, 3500, \\4000, 4500, \\5016\}   \end{tabular}                          \\
Ground station range                                  &~  & km                      & 1123                                                                   \\
Node delay                                  & $\textit{T}_{node}$  & ms                      & 10                                                                   \\
Speed of light                              & $\textit{c}_s$      & m/s                     & 299792458                                                          \\
\begin{tabular}[c]{@{}l@{}}Satellite transmission power \\ constraint\end{tabular}           & $\textit{P}_{lim}$      & ~                     &\begin{tabular}[c]{@{}l@{}}\{0.1 W, 0.3 W,\\ 0.5 W\}\end{tabular}\\ 
Node-degree	 	& ~               &~  		& 4  \\
\hline
\end{tabular}
\arrayrulecolor{black}
\end{table}

\subsection{Network Latency Without Transmission Power Constraints}

We first investigate network latency without any constraints on satellite transmission power using the formulation in (22)–(25). Table \ref{table2} shows the network latency for all five inter-continental connections and the path for each inter-continental connection at a 1575 km LISL range at the first time slot. At minimum feasible LISL range, the total network latency for all five inter-continental connections is 594.9 ms at the first time slot. Table \ref{table3} presents the network latency and paths for all inter-continental connections for the 5016 km LISL range at the first time slot. Compared to the results in Table \ref{table2}, the network latency for each inter-continental connection as well as the total network latency is smaller, and the number of hops/satellites on the path also decreases at a LISL range of 5016 km. The total network latency for all five inter-continental connections is 310.00 ms at this LISL range.
\begin{table*}
\centering
\caption{Network Latency, Path, and Number of Hops for 1575 km LISL Range at the First Time Slot Without Transmission Power Constraints. }
\label{table2}
\setlength{\tabcolsep}{0.7pt}
\arrayrulecolor{black}
\begin{tabular}{!{\color{black}\vrule}c!{\color{black}\vrule}c!{\color{black}\vrule}c!{\color{black}\vrule}c!{\color{black}\vrule}} 
\hline
Inter-Continental Connection & \begin{tabular}[c]{@{}c@{}}Network \\Latency\textit{ }(ms)\end{tabular} & Path                                                                                                                                                & Number
  of Hops  \\ 
\hline
New
  York - London          & 71.12                                                                   & \begin{tabular}[c]{@{}c@{}}GS at New York, satellites 11872, 11511, 11805, 11807, \\GS at London\end{tabular}                                                & 4                 \\ 
\hline
Cairo
  – Tokyo              & 105.20                                                                  & \begin{tabular}[c]{@{}c@{}}GS at Cairo, satellites 10310, 10311, 12108, 10413, 10512, \\10223, 10807, GS at Tokyo\end{tabular}                               & 7                 \\ 
\hline
Sao
  Paulo - Istanbul       & 129.91                                                                  & \begin{tabular}[c]{@{}c@{}}GS at Sao Paulo, satellites 10168, 10169, 10171, 10101, 11326, \\11423, 11520, 10208, 10209, GS at Istanbul\end{tabular}          & 9                 \\ 
\hline
Cape
  Town - Sydney         & 131.12                                                                  & \begin{tabular}[c]{@{}c@{}}GS at Cape Town, satellites 11339, 10953, 11344, 11346, 11543, \\11448, 11450, 11452, 11453, GS at Sydney\end{tabular}            & 9                 \\ 
\hline
Mexico
  City - Shanghai     & 157.53                                                                  & \begin{tabular}[c]{@{}c@{}}GS at Mexico City, satellites 10923, 10921, 10919, 10917, 11109, \\10912, 10618, 10422, 10324, 10226, GS at Shanghai\end{tabular} & 10                \\
\hline
\end{tabular}
\arrayrulecolor{black}
\end{table*}

\begin{table*}
\centering
\caption{Network Latency, Path, and Number of Hops at 5016 km LISL at the First Time Slot Without Transmission Power Constraints. }
\label{table3}
\setlength{\tabcolsep}{0.7pt}
\arrayrulecolor{black}
\begin{tabular}{!{\color{black}\vrule}c!{\color{black}\vrule}c!{\color{black}\vrule}c!{\color{black}\vrule}c!{\color{black}\vrule}} 
\hline
Inter-Continental
  Connection & \begin{tabular}[c]{@{}c@{}}Network \\Latency\textit{ }(ms)\end{tabular} & Path                                                       & Number
  of Hops  \\ 
\hline
New
  York - London            & 40.93                                                                   & GS at New York, satellites 11872, 11807, GS at
  London             & 2                 \\ 
\hline
Cairo
  - Tokyo                & 54.02                                                                   & GS at Cairo, satellites 12010, 10223, GS at Tokyo                   & 2                 \\ 
\hline
Sao
  Paulo - Istanbul         & 68.84                                                                   & GS at Sao Paulo, satellites 10168, 11326, 11617,
  GS at Istanbul   & 3                 \\ 
\hline
Cape
  Town - Sydney           & 70.12                                                                   & GS at Cape Town, satellites 11341, 11544, 11453,
  GS at Sydney     & 3                 \\ 
\hline
Mexico
  City - Shanghai       & 76.08                                                                   & GS at Mexico City, satellites 11670, 10816, 10324,
  GS at Shanghai & 3                 \\
\hline
\end{tabular}
\arrayrulecolor{black}
\end{table*}

\begin{figure*}[htbp]
\centerline{\includegraphics[width=16cm, height=7cm]{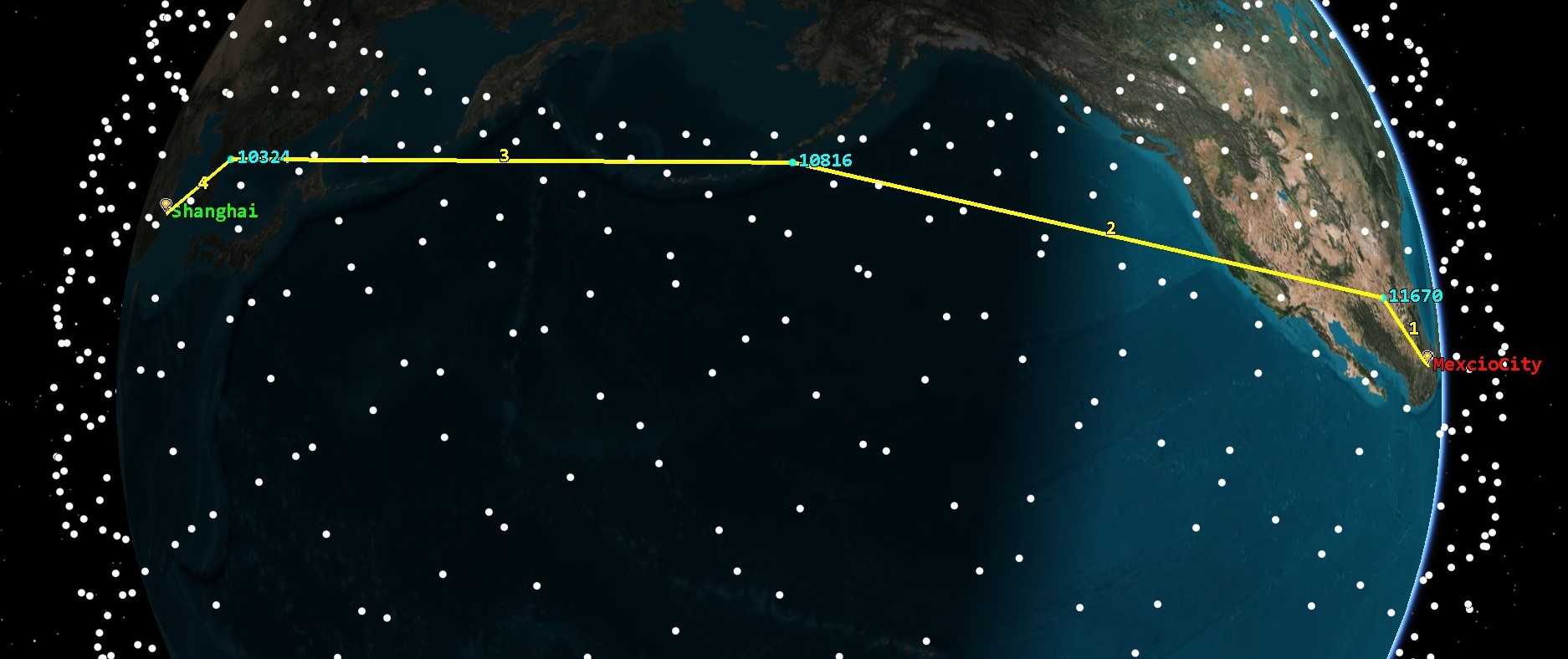}}
\caption{An illustration of the path at 5016 km LISL range without transmission power constraints for the inter-continental connection from Mexico City to Shanghai at the first time slot.}
\label{figure1}
\end{figure*}

\begin{figure}[htbp]
\centerline{\includegraphics[width=0.4788\textwidth]{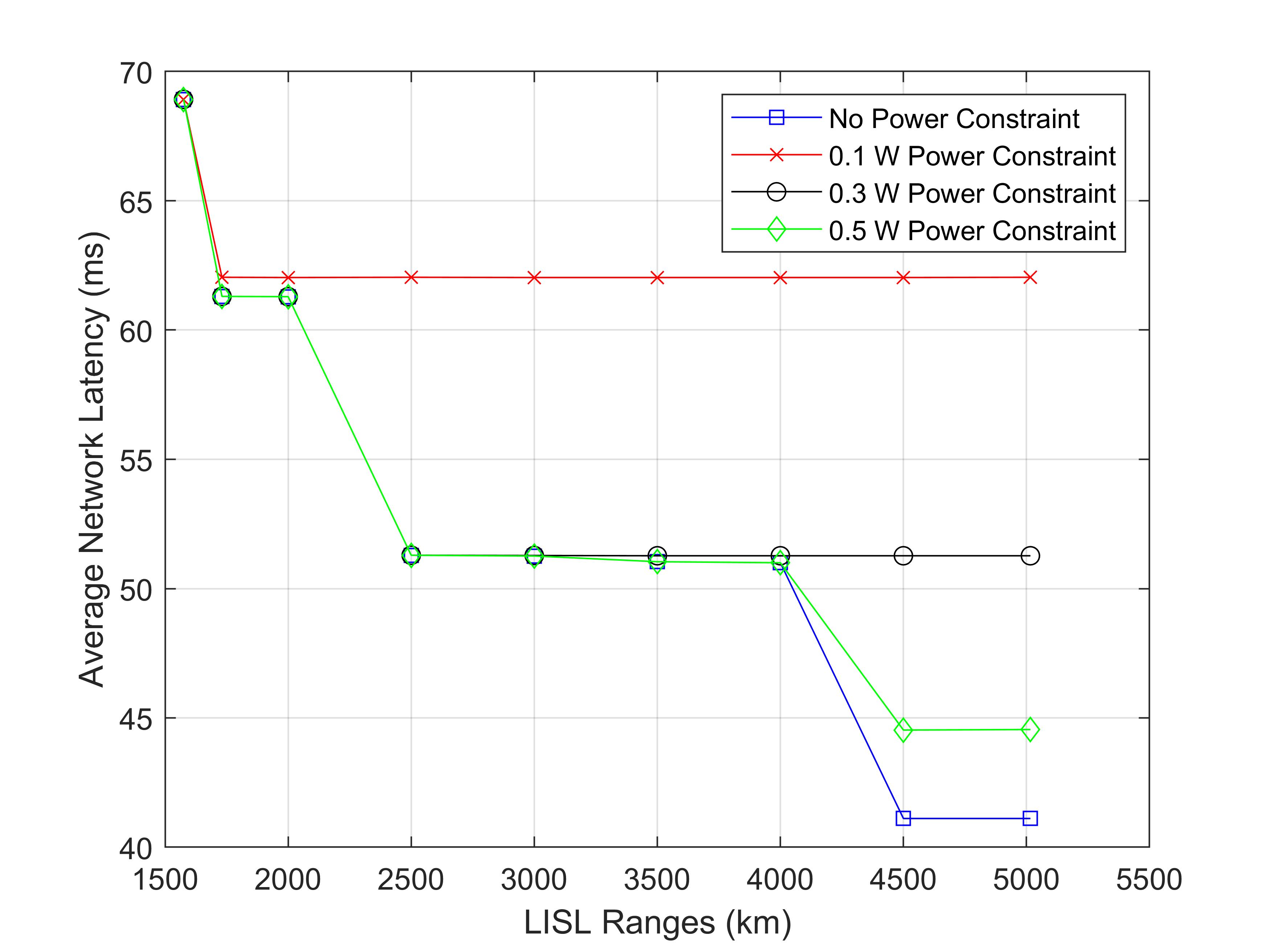}}
\caption{Average network latency for different LISL ranges with and without power constraints for inter-continental connection from New York to London over the first 100 time slots. }
\label{figure2}
\end{figure}

\begin{figure}[htbp]
\centerline{\includegraphics[width=0.4788\textwidth]{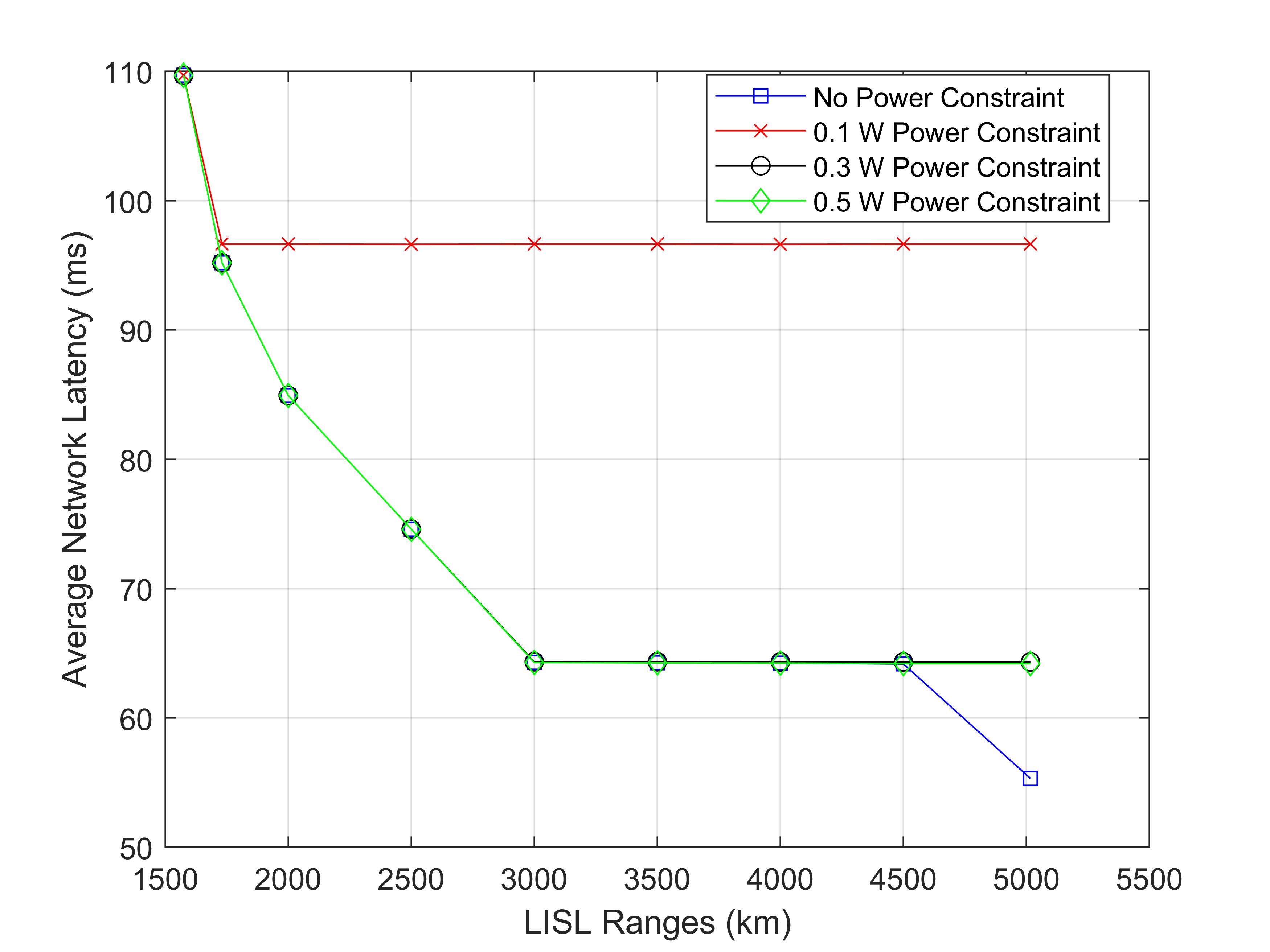}}
\caption{Average network latency for different LISL ranges with and without power constraints for inter-continental connection from Cairo to Tokyo over the first 100 time slots.}
\label{figure3}
\end{figure}

\begin{figure}[htbp]
\centerline{\includegraphics[width=0.4788\textwidth]{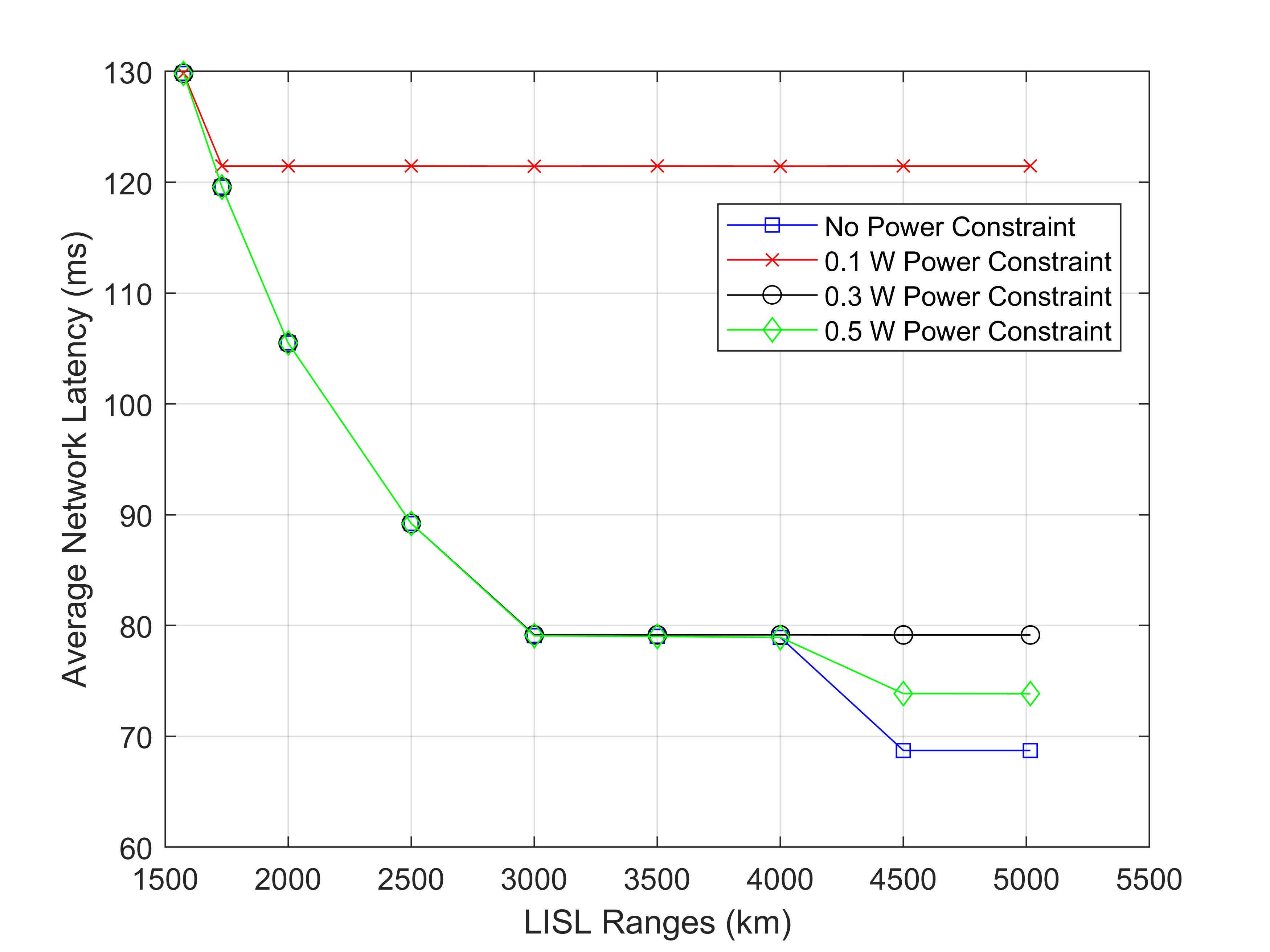}}
\caption{Average network latency for different LISL ranges with and without power constraints for inter-continental connection from Sao Paulo to Istanbul over the first 100 time slots.}
\label{figure4}
\end{figure}

\begin{figure}[htbp]
\centerline{\includegraphics[width=0.4788\textwidth]{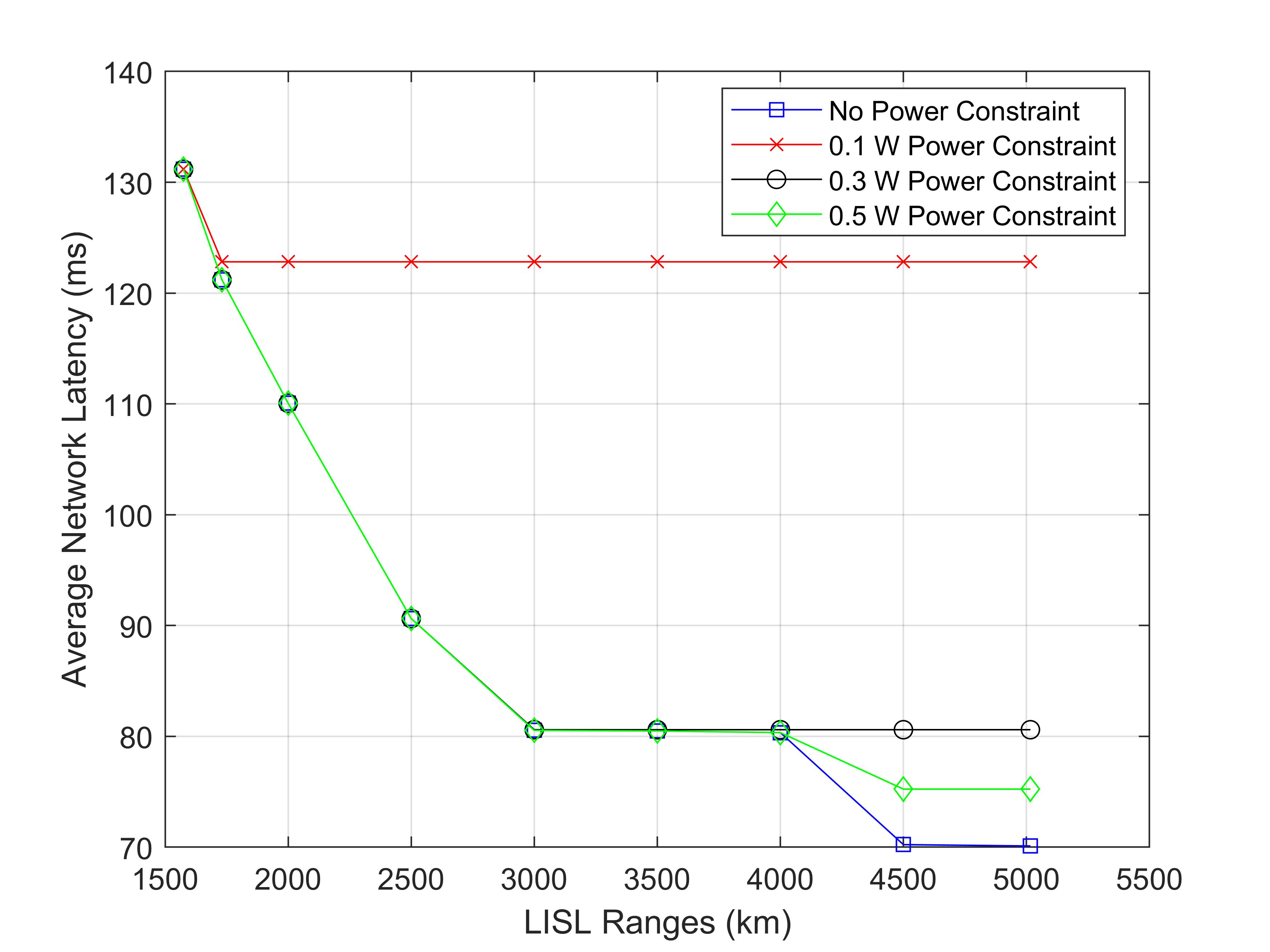}}
\caption{Average network latency for different LISL ranges with and without power constraints for inter-continental connection from Cape Town to Sydney over the first 100 time slots.}
\label{figure5}
\end{figure}

\begin{figure}[htbp]
\centerline{\includegraphics[width=0.4788\textwidth]{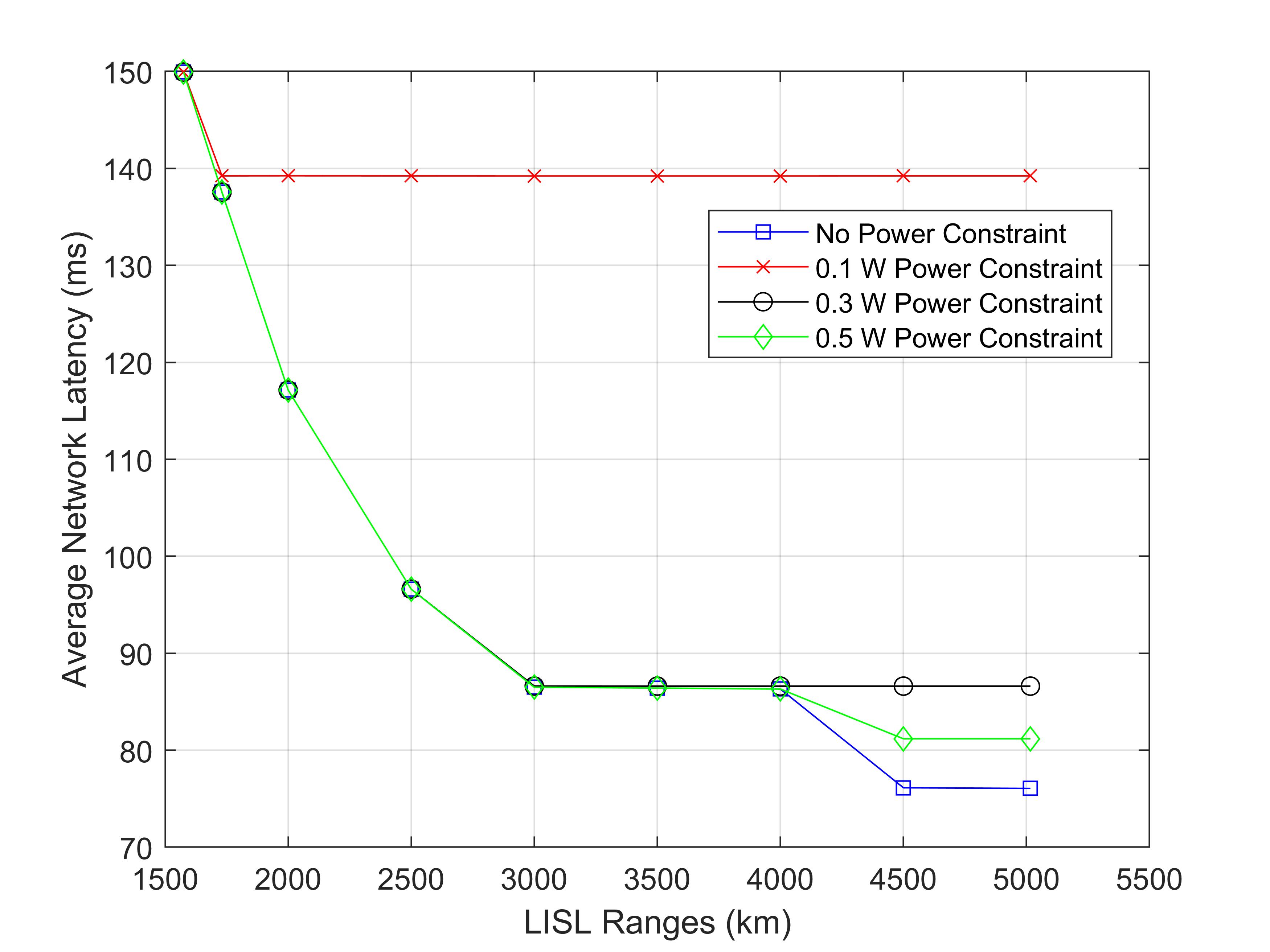}}
\caption{Average network latency for different LISL ranges with and without power constraints for inter-continental connection from Mexico City to Shanghai over the first 100 time slots.}
\label{figure6}
\end{figure}

\begin{figure}[htbp]
\centerline{\includegraphics[width=0.4788\textwidth]{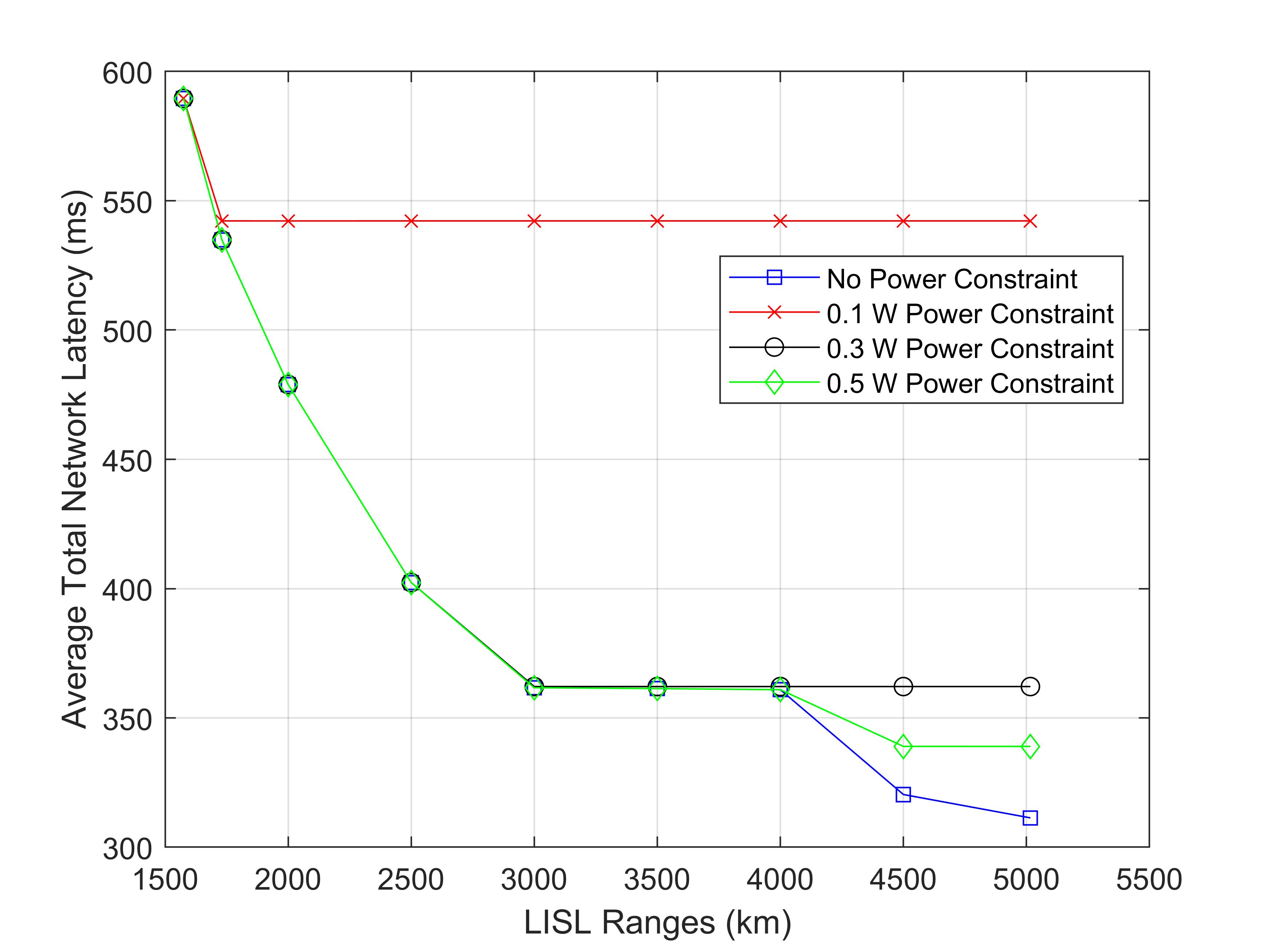}}
\caption{Average total network latency for different LISL ranges with and without power constraints for all five inter-continental connections over the first 100 time slots. }
\label{figure7}
\end{figure}

Figure \ref{figure1} illustrates the 5016 km LISL range path for inter-continental connection from Mexico City to Shanghai without power constraints. This path comprises three satellites and four optical links. As depicted in the figure, for this extensive inter-continental connection with LISL range as 5016 km, we employ satellites 11670, 10816, and 10324 from the Starlink Phase 1 Version 3 constellation. It's important to note that at this maximum LISL range, satellites cannot establish a line-of-sight optical link when the distance between two satellites exceeds 5016 km. Conversely, when selecting the shortest path, a satellite will opt to establish an optical link with the farthest one within its maximum LISL range.
Figures \ref{figure2} to \ref{figure6} show the average network latency for each inter-continental connection, i.e., the average of the network latency for each inter-continental connection over 100 time slots, at all LISL ranges. Figure \ref{figure7} shows the average total network latency for all five inter-continental connections, i.e., the average of the total network latency for all five inter-continental connections over 100 time slots, at all LISL ranges. There is a clear trend, as indicated by the blue curve with square markers: with the increase in LISL range, the average network latency and the average total network latency decrease when there is no limit on the satellite transmission power. The highest average network latency and average total network latency occur at the minimum feasible LISL range of 1575 km, while the lowest occurs at the maximum feasible LISL range of 5016 km. This happens because satellites can establish longer ISLs at the longer LISL range, and hence fewer ISLs and satellites are required for an inter-continental connection, so less node delays are added to the network latency for a given inter-continental connection.

\subsection{Network Latency with Transmission Power Constraints}
We further examine network latency with satellite transmission power constraints using the formulation in (22)–(25). We initially assume a satellite transmission power limit of 0.3 W. Table \ref{table4} shows the network latency for all five inter-continental connections and the path for each inter-continental connection at a 1575 km LISL range, at the first time slot. When the LISL range has this minimum feasible value, the total network latency for all five inter-continental connections is 594.9 ms at the first time slot. Table \ref{table5} shows the network latency for all inter-continental connections and the path for each inter-continental connection for the 5016 km LISL range at the first time slot. When comparing the results in Table \ref{table4} and Table \ref{table5}, it becomes evident that at the 5016 km LISL range, the network latency for each intercontinental connection, as well as the total network latency, is reduced, and there is also a decrease in the number of hops/satellites on the path. The total network latency for the five inter-continental connections is 361.36 ms.

Table \ref{table4} shows that with a 0.3 W satellite transmission power constraint, the network latency for all inter-continental connections is the same as that without satellite transmission power constraints in Table \ref{table2} at the 1575 km LISL range. This is because at this LISL range, satellites are not able to establish optical links which consume transmission power higher than this limit. When comparing results in Tables \ref{table5} and \ref{table3}, we find that after adding the 0.3 W satellite transmission power constraints, the network latency for all inter-continental connections increases significantly at the 5016 km LISL range, and the total network latency for all inter-continental connections increases by about 50 ms. 
\begin{table*}
\centering
\caption{Network Latency, Path, and Number of Hops for a 1575 km LISL Range at the First Time Slot With 0.3 W Transmission Power Constraint.}
\label{table4}
\setlength{\tabcolsep}{0.7pt}
\arrayrulecolor{black}
\begin{tabular}{!{\color{black}\vrule}c!{\color{black}\vrule}c!{\color{black}\vrule}c!{\color{black}\vrule}c!{\color{black}\vrule}} 
\hline
Inter-Continental Connection & \begin{tabular}[c]{@{}c@{}}Network \\Latency (ms)\end{tabular} & 
  Path                                                                                                                                              & Number
  of Hops  \\ 
\hline
New
  York - London          & \begin{tabular}[c]{@{}c@{}}71.12 \\~\end{tabular}              & \begin{tabular}[c]{@{}c@{}}GS at New York, satellites 11872, 11511, 11805, 11807, \\GS at London\end{tabular}                                                & 4                 \\ 
\hline
Cairo
  – Tokyo              & \begin{tabular}[c]{@{}c@{}}105.20 \\~\end{tabular}             & \begin{tabular}[c]{@{}c@{}}GS at Cairo, satellites 10310, 10311, 12108, 10413, 10512, \\10223, 10807, GS at Tokyo\end{tabular}                               & 7                 \\ 
\hline
Sao
  Paulo - Istanbul       & \begin{tabular}[c]{@{}c@{}}129.91 \\~\end{tabular}             & \begin{tabular}[c]{@{}c@{}}GS at Sao Paulo, satellites 10168, 10169, 10171, 10101, 11326, \\11423, 11520, 10208, 10209, GS at Istanbul\end{tabular}          & 9                 \\ 
\hline
Cape
  Town - Sydney         & \begin{tabular}[c]{@{}c@{}}131.12 \\~\end{tabular}             & \begin{tabular}[c]{@{}c@{}}GS at Cape Town, satellites 11339, 10953, 11344, 11346, 11543, \\11448, 11450, 11452, 11453, GS at Sydney\end{tabular}            & 9                 \\ 
\hline
Mexico
  City - Shanghai     & \begin{tabular}[c]{@{}c@{}}157.53 \\~\end{tabular}             & \begin{tabular}[c]{@{}c@{}}GS at Mexico City, satellites 10923, 10921, 10919, 10917, 11109, \\10912, 10618, 10422, 10324, 10226, GS at Shanghai\end{tabular} & 10                \\
\hline
\end{tabular}
\arrayrulecolor{black}
\end{table*}

\begin{figure*}[htbp]
\centerline{\includegraphics[width=16cm, height=7cm]{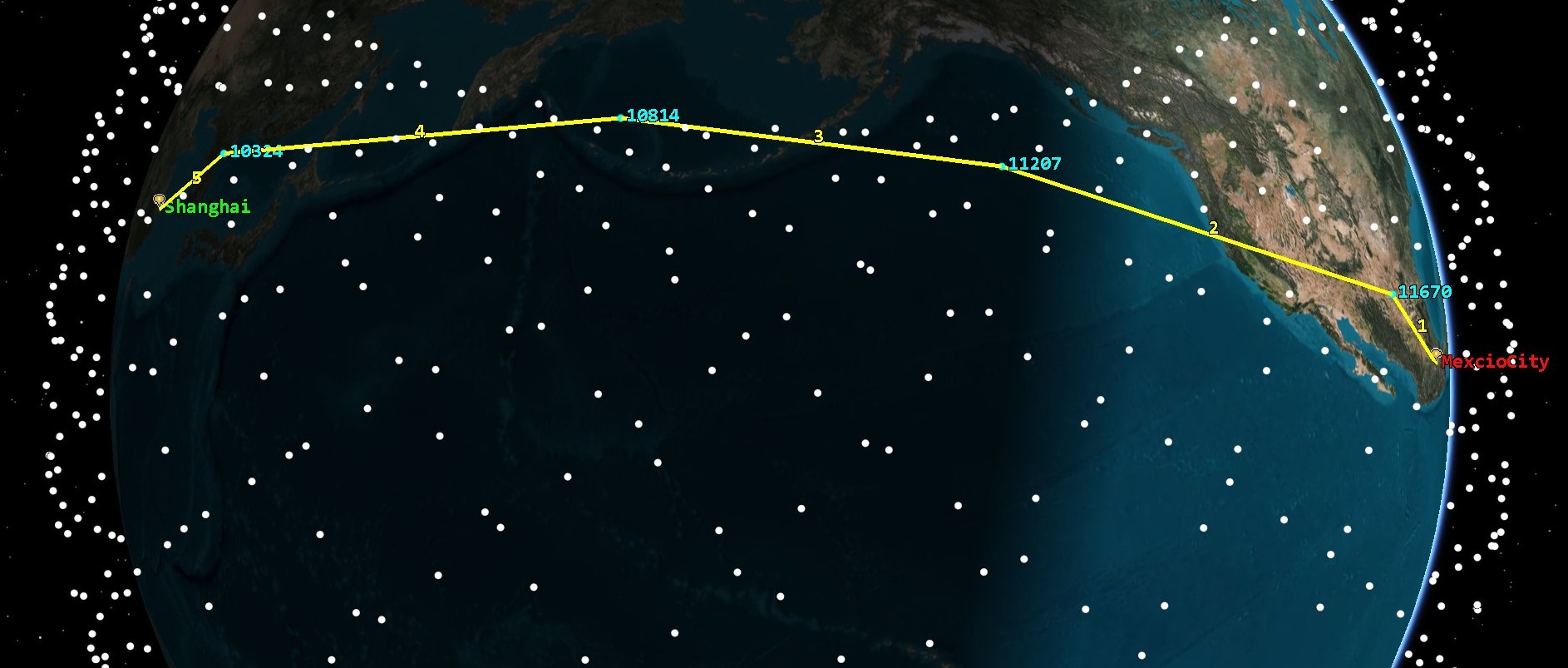}}
\caption{An illustration of the path at the 5016 km LISL range with 0.3 W transmission power constraints for the inter-continental connection from Mexico City to Shanghai at the first time slot. }
\label{figure8}
\end{figure*}

\begin{table*}
\centering
\caption{Network Latency, Path, and Number of Hops for 5016 km LISL Range at the First Time Slot With 0.3 W Transmission Power Constraints.}
\label{table5}
\setlength{\tabcolsep}{0.7pt}
\arrayrulecolor{black}
\begin{tabular}{!{\color{black}\vrule}c!{\color{black}\vrule}c!{\color{black}\vrule}c!{\color{black}\vrule}c!{\color{black}\vrule}} 
\hline
Inter-Continental Connection & \begin{tabular}[c]{@{}c@{}}Network \\Latency (ms)\end{tabular} & 
  Path                                                          & Number
  of Hops  \\ 
\hline
New
  York - London          & 51.05                  & GS at New York, satellites 11872, 11413, 11807, GS at London             & 3                 \\ 
\hline
Cairo
  - Tokyo              & 64.25                  & GS at Cairo, satellites 12010, 10512, 10223, GS at Tokyo                 & 3                 \\ 
\hline
Sao
  Paulo - Istanbul       & 79.16                  & GS at Sao Paulo, satellites 10168, 11229, 11423, 11617, GS at Istanbul   & 4                 \\ 
\hline
Cape
  Town - Sydney         & 80.49                  & GS at Cape Town, satellites 11341, 11345, 11642, 11453, GS at Sydney     & 4                 \\ 
\hline
Mexico
  City - Shanghai     & 86.41                  & GS at Mexico City, satellites 11670, 11207, 10814, 10324, GS at Shanghai & 4                 \\
\hline
\end{tabular}
\arrayrulecolor{black}
\end{table*}
Figure \ref{figure8} illustrates the path at the 5016 km LISL range for the inter-continental connection from Mexico City to Shanghai with a 0.3 W transmission power constraint. Figures \ref{figure1} and \ref{figure8} show that there are fewer links and nodes on the path in Figure \ref{figure1}, and therefore less network latency when there is no constraint on transmission power. On the other hand, the links on the path, shown in Figure \ref{figure1}, are longer, so the satellites on this path use more transmission power to establish these links. 
From Figures \ref{figure2} to \ref{figure7}, we see that after adding a 0.3 W transmission power constraint, the average network latency for each inter-continental connection and the average total network latency for all inter-continental connections level off at a LISL range of 3000 km, as indicated by the black curve with circle markers. The latencies are the same after this LISL range as longer optical links cannot exist with this constraint on satellite optical link transmission power. We also investigate the average network latency and average total network latency with different transmission power constraints, including 0.1 W and 0.5 W. From Figures \ref{figure2} to \ref{figure7}, we find that with 0.1 W and 0.5 W transmission power constraints, as indicated by the red curve with X markers and green curve with diamond markers, respectively, the average network latency and average total network latency stop changing when the LISL range is greater than 1731 km and 4500 km, respectively. 

With different transmission power constraints, we find that the tighter the constraint on satellite transmission power, the smaller the LISL range when the average network latency or average total network latency stops decreasing. For smaller values of satellite optical link transmission power constraints, the limitation on the optical link distance is tighter. Consequently, only shorter optical links can be established, and thereby more optical links and nodes are required to establish an inter-continental connection regardless of the increase in LISL range.

\begin{figure}[htbp]
\centerline{\includegraphics[width=0.4788\textwidth]{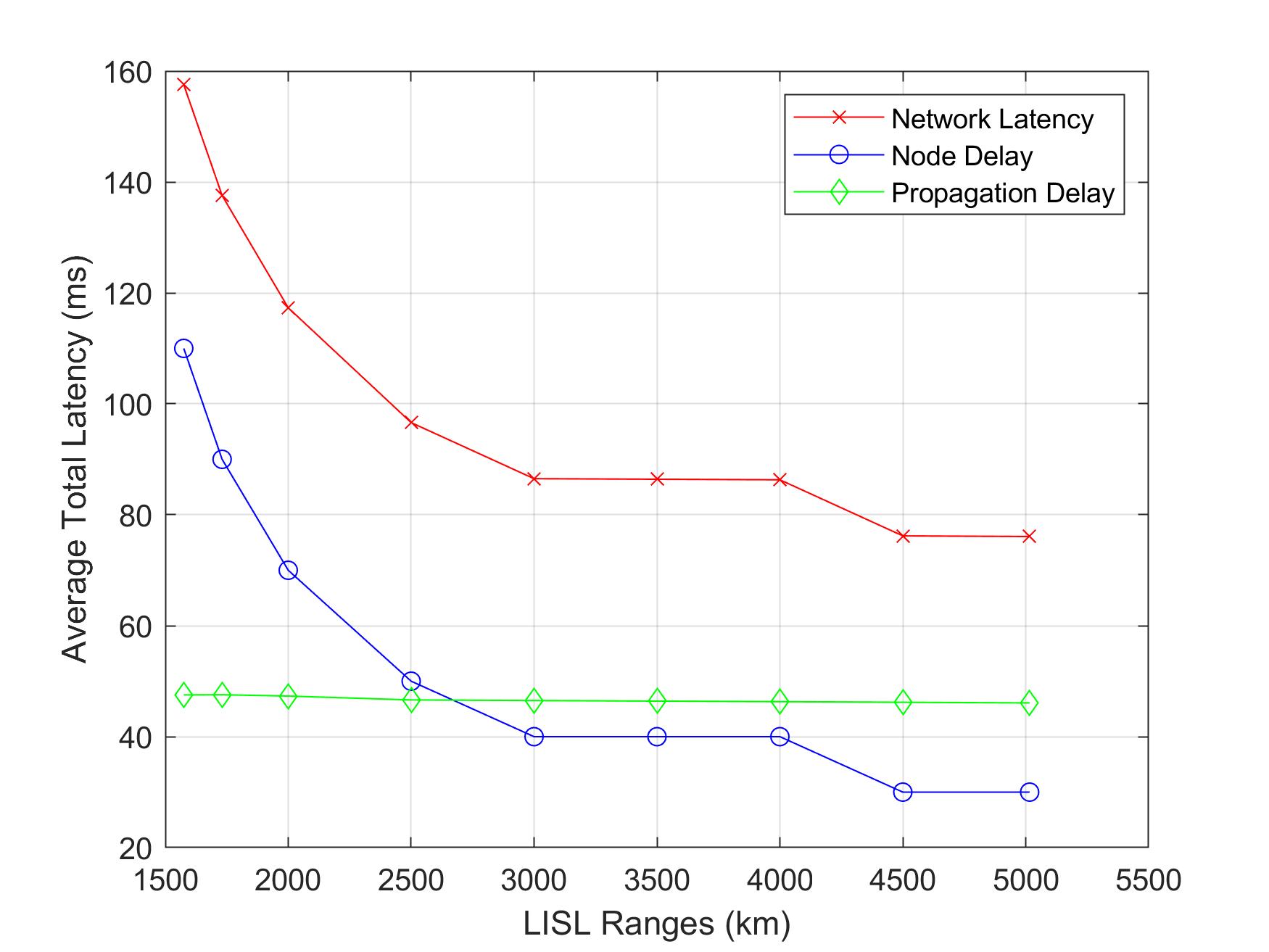}}
\caption{Average total network latency, node delay and propagation delay for different LISL ranges without power constraints for inter-continental connection from Mexico City to Shanghai over the first 100 time slots. }
\label{figure9}
\end{figure}

To investigate the specific relationship and variations in propagation delays and node delays concerning changes in the LISL range, we have constructed Figure \ref{figure9}. This figure illustrates the curves depicting the average total network latency, average total node delay, and average total propagation delay for different LISL ranges, all without power constraints. These results pertain to the inter-continental connection from Mexico City to Shanghai over the first 100 time slots. From the figure, we observe that as the LISL range increases, several trends become evident. Firstly, the average total propagation delay of the path exhibits a very slight decrease. Secondly, the average total node delay of the path has a noticeable reduction. Finally, the average total network latency of the path shows a decline as well.

\subsection{Insights}
There are two insights derived from our analysis that we think will be helpful in practice for optical satellite communications:

\begin{itemize}
    \item The network latency of the path for an inter-continental connection between two ground stations decreases with an increase in LISL range when there is no constraint on satellite transmission power. When LISL range increases, there are longer links on the path and the propagation distance of an optical link increases, fewer links and nodes are needed to establish the path, total propagation delay of the path decreases, total node delay of the path decreases, and network latency of the path decreases. For example, at a 2000 km LISL range, network latency for the New York - London inter-continental connection is 61.29 ms at the first time slot, while for a 3000 km LISL range it decreases to 51.26 ms.

   \item Under satellite transmission power constraints, the average network latency doesn’t decrease with the increase in LISL range beyond a certain LISL range and levels off at that LISL range. The transmission power constraints limit the propagation distance of optical links, and they cannot exceed this propagation distance limitation regardless of the increase in LISL range. For tighter transmission power constraints, the propagation distance limitation is tighter, and thereby the increase in LISL range stops affecting the average network latency at smaller LISL range values. For instance, for the New York-London inter-continental connection under 0.1 W transmission power constraints, the average network latency over 100 time slots remains around 62 ms for LISL ranges longer than 1731 km, while for 0.5 W transmission power constraints, it stays at approximately 44 ms for LISL ranges longer than 4500 km.

\end{itemize}

\section{Conclusions}
We propose a binary integer linear programming formulation to minimize total network latency under various realistic constraints, including satellite transmission power constraints, in the FSOSN resulting from the Starlink Phase 1 Version 3 constellation. This formulation provides more accurate results while also being able to handle multiple inter-continental connections simultaneously, compared to earlier works evaluating the trade-off between network latency and satellite transmission power. Solution times are much longer than in our previous work: about 7 hours for the maximum 5016 km LISL range for the Starlink Phase 1 Version 3 constellation. This formulation provides accurate results but is unsuitable for real-time solution of real-world problems which may have hundreds of simultaneous inter-continental connections. However it will help us to evaluate heuristics in our future work.

We plan to study various options for improving the solution speed, including providing an initial heuristic solution to the solver, terminating the branch and bound process early, etc.

Our formulation for minimizing network latency under satellite transmission power constraints uses appropriate system models for network latency and satellite transmission power. We examine the network latency of an inter-continental connection at a time slot, total network latency of all inter-continental connections at a time slot, average network latency of an inter-continental connection over 100 time slots, and average total network latency over 100 time slots in this FSOSN for five different inter-continental connections under nine different LISL ranges and three different values of the satellite transmission power constraints. 
 
The results in Tables 2 and 3 show that for total network latency without  satellite transmission power constraints, a longer LISL range gives better values of total network latency, and the minimum total network latency occurs when the LISL range takes the maximum feasible value of 5016 km for this constellation. With different values of satellite transmission power constraints, the average total network latency levels off at different LISL ranges, i.e., after these LISL ranges, no changes occur in the average total network latency with an increase in LISL range, and different limits on transmission power affect the average total network latency differently. With a limitation of 0.5 W, 0.3 W, and 0.1 W on the satellite transmission power, the average total network latency levels off at 339 ms, 361 ms, and 542 ms, respectively, at 4500 km, 3000 km, and 1731 km LISL ranges, respectively.

In future, we will study the problem of minimizing satellite transmission power under various realistic constraints, including network latency. We will also investigate the problem of finding an appropriate LISL range for balancing total network latency and satellite transmission power in an FSOSN by formulating this problem as a multi-objective mathematical program. In future research, we will also study the problem that when to switch between LEO satellite network and terrestrial network based on latency and power constraints, which are critical for LEO satellite network. Furthermore, it's essential to note that satellite transmission power represents just one facet of the overall satellite link budget. Satellites also have inherent limitations on onboard power consumption, which consequently restrict their processing and routing capabilities. In our future research, we plan to delve into the impact of onboard satellite power consumption on network latency.We assume that the router possesses sufficient processing capabilities, and we impose link-disjoint constraints, ensuring that all flows do not exceed one connection in this work. In other words, if a link is already in use by a particular source-destination route, other source-destination pairs must seek alternative links for transmitting their data. In future research, we aim to explore the problems of minimizing network latency in scenarios where an optical link can be utilized by multiple flows from various inter-continental connections.

\section*{Acknowledgement}
This work was supported by the High Throughput and Secure Networks Challenge Program at the National Research Council of Canada.

\begin{IEEEbiography}[{\includegraphics[width=1in,height=1.25in,clip,keepaspectratio]{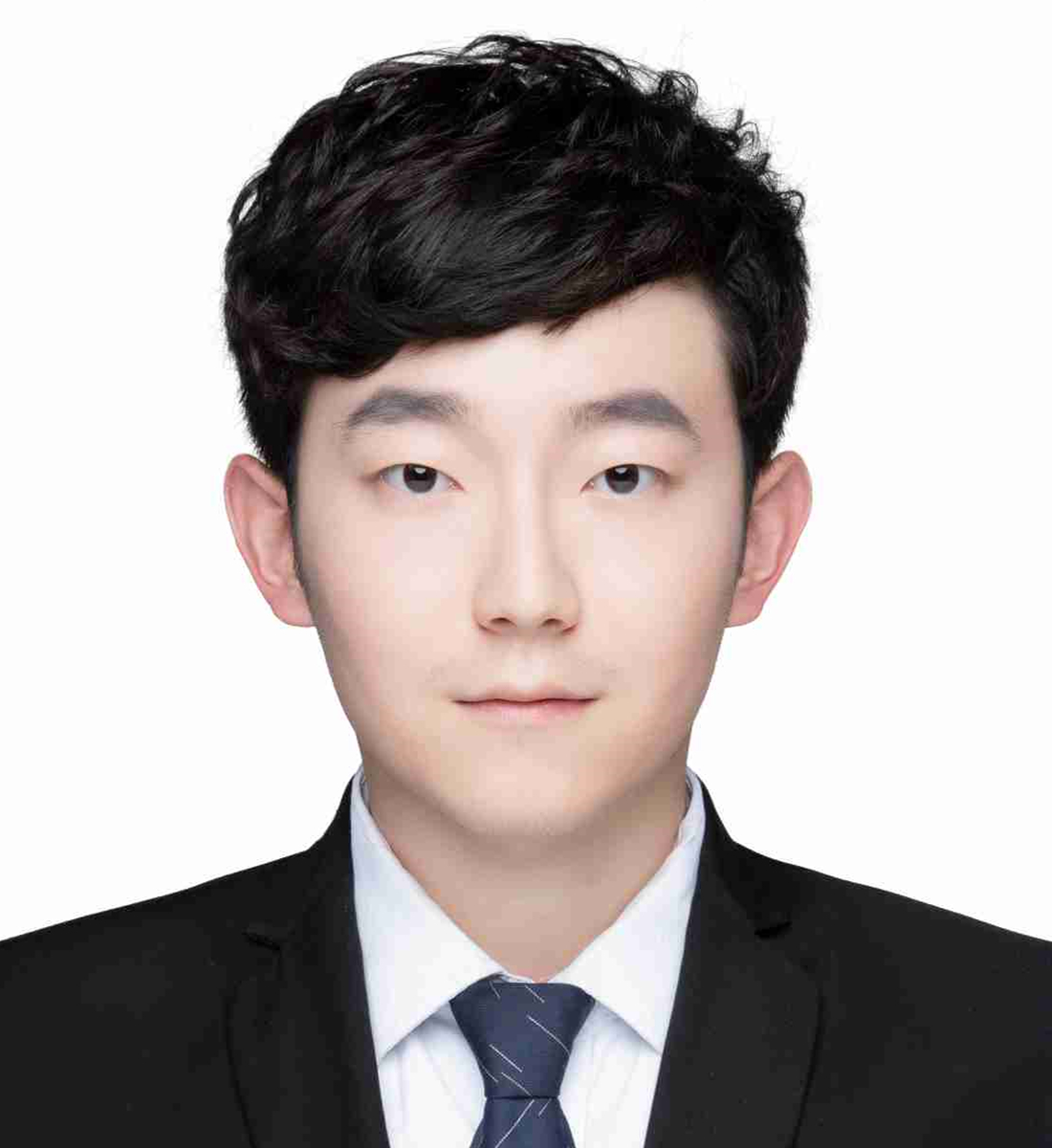}}]{Jintao Liang}
\:(Member, IEEE) received his Bachelor of Engineering degree in Automation Engineering from Shandong University (China) in 2019, and his M.A.Sc. degree in the Department of Systems and Computer Engineering in Carleton University (Canada) in 2023. \par
He is currently a Phd Student in the Department of Systems and Computer Engineering in Carleton University (Canada) since 2023. \par
Mr. Liang's research interest includes free-space optical satellite networks.\par
\end{IEEEbiography}

\begin{IEEEbiography}[{\includegraphics[width=1in,height=1.25in,clip,keepaspectratio]{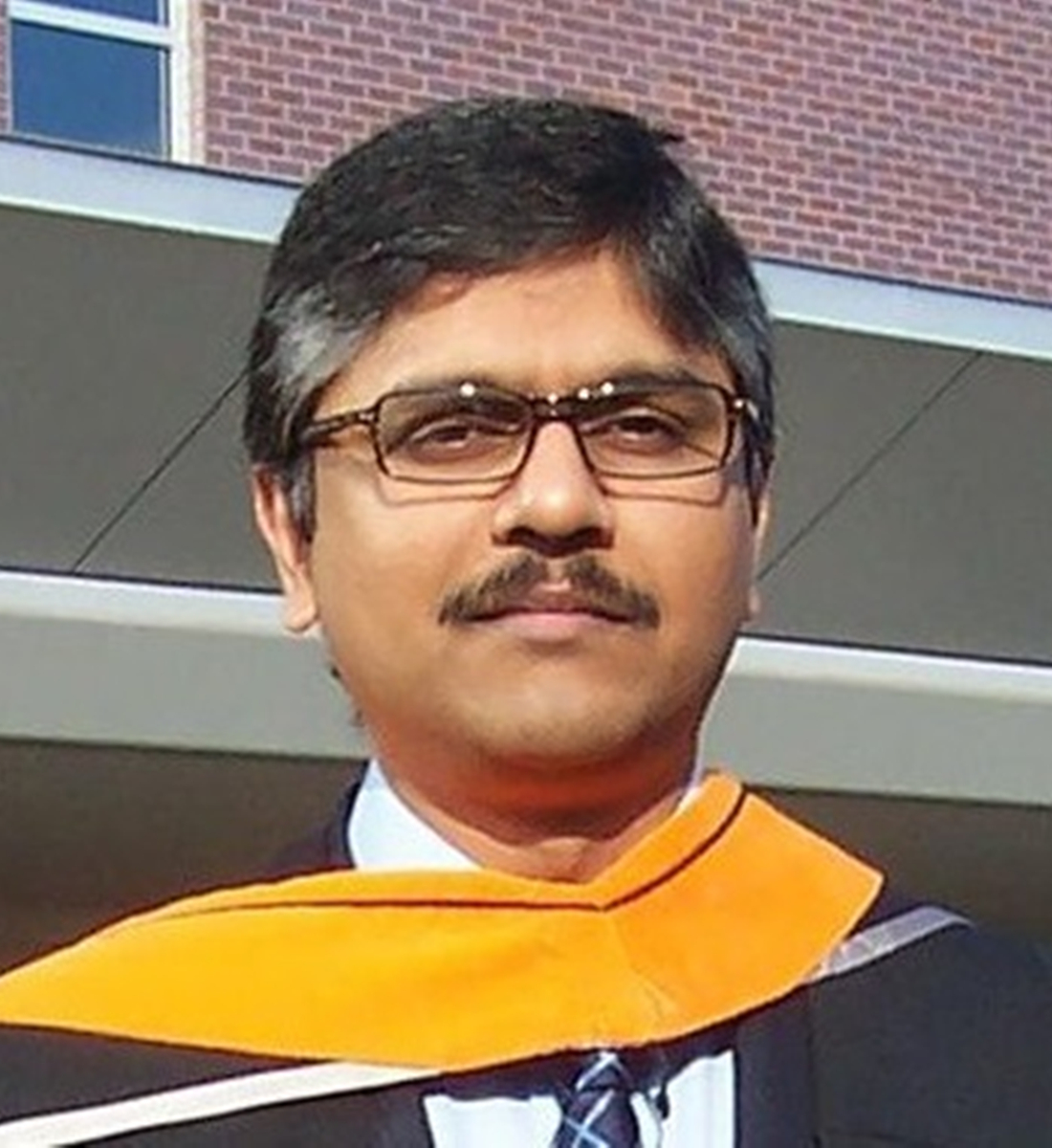}}]{Aizaz U. Chaudhry}
\:(Senior Member, IEEE) received his B.Sc. degree in Electrical Engineering from the University of Engineering and Technology Lahore (Pakistan) in 1999, and his M.A.Sc. and Ph.D. degrees in Electrical and Computer Engineering from Carleton University (Canada) in 2010 and 2015, respectively.\par
He is currently a Senior Research Associate with the Department of Systems and Computer Engineering at Carleton University. Previously, he worked as an NSERC Postdoctoral Research Fellow at Communications Research Centre Canada, Ottawa. His research work has been published in refereed venues, and has received several citations. He has authored and co-authored more than thirty-five publications. His research interests include the application of machine learning and optimization in wireless networks.\par
Dr. Chaudhry is a licensed Professional Engineer in the Province of Ontario, a Senior Member of IEEE, and a Member of IEEE ComSoc’s Technical Committee on Satellite and Space Communications. He serves as a technical reviewer for conferences and journals on a regular basis. He has also served as a TPC Member for conferences, such as IEEE ICC 2021 Workshop 6GSatComNet, IEEE ICC 2022 Workshop 6GSatComNet, IEEE VTC 2022-Spring, IEEE VTC 2022-Fall, the twenty-first international conference on networks (ICN 2022), eleventh advanced satellite multimedia systems conference (ASMS 2022), seventeenth signal processing for space communications workshop (SPSC 2022), and the first international symposium on satellite communication systems and services (SCSS 2022).\par
\end{IEEEbiography}

\begin{IEEEbiography}[{\includegraphics[width=1in,height=1.25in,clip,keepaspectratio]{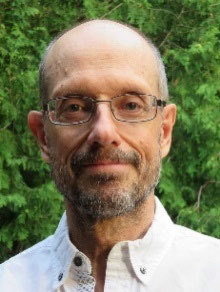}}]{John W. Chinneck}
\: received the Ph.D. degree in systems design engineering from the University of Waterloo, in 1983. \par
He researches the development and application of optimization algorithms. He served as the Editor-in-Chief for the INFORMS Journal on Computing from 2007 to 2012, and the Chair for the INFORMS Computing Society from 2006 to 2007. He has received the Research AchievementAward from Carleton University, and the Award of Merit from the Canadian Operational Research Society, among other awards.\par
\end{IEEEbiography}

\begin{IEEEbiography}[{\includegraphics[width=1in,height=1.25in,clip,keepaspectratio]{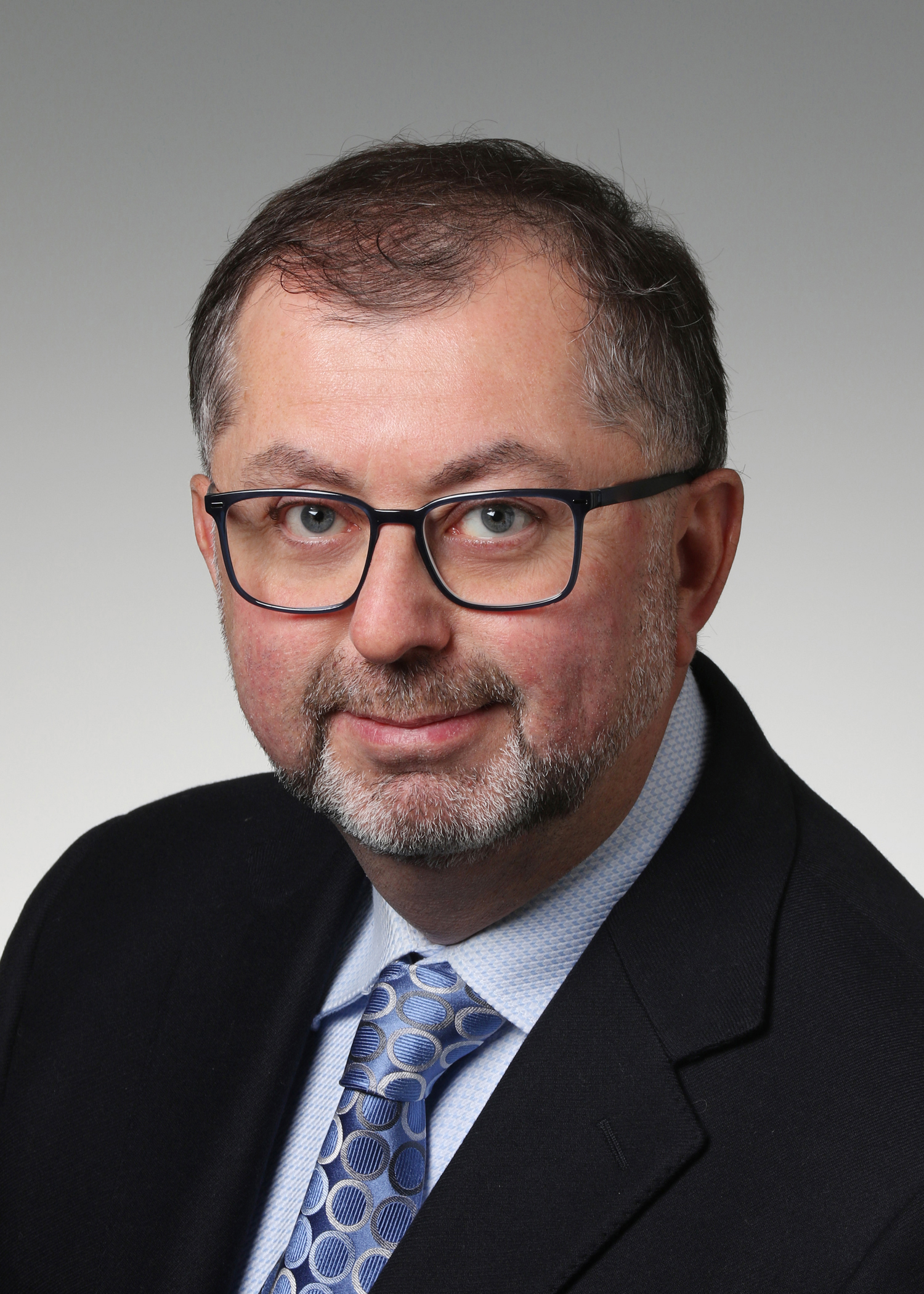}}]{Halim Yanikomeroglu}
\:(Fellow, IEEE)received the BSc degree in electrical and electronics engineering from the Middle East Technical University, Ankara, Turkey, in 1990, and the MASc degree in electrical engineering (now ECE) and the PhD degree in electrical and computer engineering from the University of Toronto, Canada, in 1992 and 1998, respectively. Since 1998 he has been with the Department of Systems and Computer Engineering at Carleton University, Ottawa, Canada, where he is now a Full Professor. \par

Dr. Yanikomeroglu’s research interests cover many aspects of wireless communications and networks, with a special emphasis on non-terrestrial networks (NTN) in the recent years. He has given 110+ invited seminars, keynotes, panel talks, and tutorials in the last five years. He has supervised or hosted over 150 postgraduate researchers in his lab at Carleton. Dr. Yanikomeroglu’s extensive collaborative research with industry resulted in 39 granted patents. Dr. Yanikomeroglu is a Fellow of the IEEE, the Engineering Institute of Canada (EIC), and the Canadian Academy of Engineering (CAE). He is a Distinguished Speaker for the IEEE Communications Society and the IEEE Vehicular Technology Society, and an Expert Panelist of the Council of Canadian Academies (CCA|CAC). \par

Dr. Yanikomeroglu is currently serving as the Chair of the Steering Committee of IEEE’s flagship wireless event, Wireless Communications and Networking Conference (WCNC). He is also a member of the IEEE ComSoc Governance Council, IEEE ComSoc GIMS, IEEE ComSoc Conference Council, and IEEE PIMRC Steering Committee. He served as the General Chair and Technical Program Chair of several IEEE conferences. He has also served in the editorial boards of various IEEE periodicals. \par

Dr. Yanikomeroglu received several awards for his research, teaching, and service, including the IEEE ComSoc Fred W. Ellersick Prize (2021), IEEE VTS Stuart Meyer Memorial Award (2020), and IEEE ComSoc Wireless Communications TC Recognition Award (2018). He received best paper awards at IEEE Competition on Non-Terrestrial Networks for B5G and 6G in 2022 (grand prize), IEEE ICC 2021, IEEE WISEE 2021 and 2022.\par
\end{IEEEbiography}

\begin{IEEEbiography}[{\includegraphics[width=1in,height=1.25in,clip,keepaspectratio]{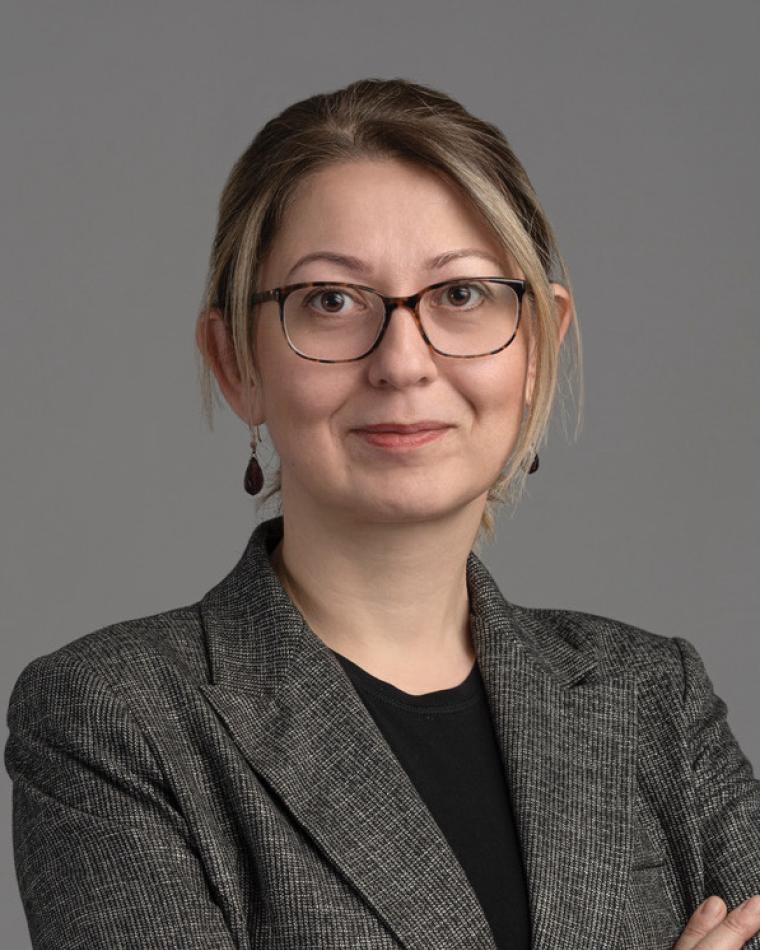}}]{Gunes Karabulut Kurt}
\:(Senior Member, IEEE) received the B.S. degree with high honors in electronics and electrical engineering from Bogazici University, Istanbul, Turkey, in 2000 and the M.A.Sc. and the Ph.D. degrees in electrical engineering from the University of Ottawa, ON, Canada, in 2002 and 2006, respectively. \par
From 2000 to 2005, she was a Research Assistant with the CASP Group, University of Ottawa. Between 2005 and 2006, she was with TenXc Wireless, Canada. From 2006 to 2008, Dr. Karabulut Kurt was with Edgewater Computer Systems Inc., Canada. From 2008 to 2010, she was with Turkcell Research and Development Applied Research and Technology, Istanbul. Between 2010 and 2021, she was with Istanbul Technical University. She is currently an Associate Professor of Electrical Engineering at Polytechnique Montréal, Montréal, QC, Canada. She is a Marie Curie Fellow and has received the Turkish Academy of Sciences Outstanding Young Scientist (TÜBA-GEBIP) Award in 2019. In addition, she is an adjunct research professor at Carleton University. \par
She is currently serving as an associate technical editor of the \textit{IEEE Communications Magazine}, an associate editor of \textit{IEEE Communication Letters}, an associate editor of \textit{IEEE Wireless Communications Letters}, and an area editor of \textit{IEEE Transactions on Machine Learning in Communications and Networking}. She is a member of the IEEE WCNC Steering Board. She is serving as the secretary of IEEE Satellite and Space Communications Technical Committee and also the chair of the IEEE special interest group entitled “Satellite Mega-constellations: Communications and Networking”. She is a Distinguished Lecturer of Vehicular Technology Society Class of 2022. \par
\end{IEEEbiography}

\begin{IEEEbiography}[{\includegraphics[width=1in,height=1.25in,clip,keepaspectratio]{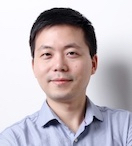}}]{Peng Hu}
\:(Senior Member, IEEE) received his Ph.D. degree in Electrical Engineering from Queen's University, Canada. \par
He is currently a Research Officer at the National Research Council Canada and an Adjunct Professor at the Cheriton School of Computer Science at the University of Waterloo. He has served as an Associate Editor of the IEEE Canadian Journal of Electrical and Computer Engineering, a voting member of the IEEE Sensors Standards committee, and on the organizing/technical committees of industry consortia and international conferences/workshops at IEEE ICC'23, IEEE PIMRC'17, IEEE AINA'15, etc. \par
His current research interests include satellite-terrestrial integrated networks, autonomous networking, and industrial Internet of Things systems. \par
\end{IEEEbiography}

\begin{IEEEbiography}[{\includegraphics[width=1in,height=1.25in,clip,keepaspectratio]{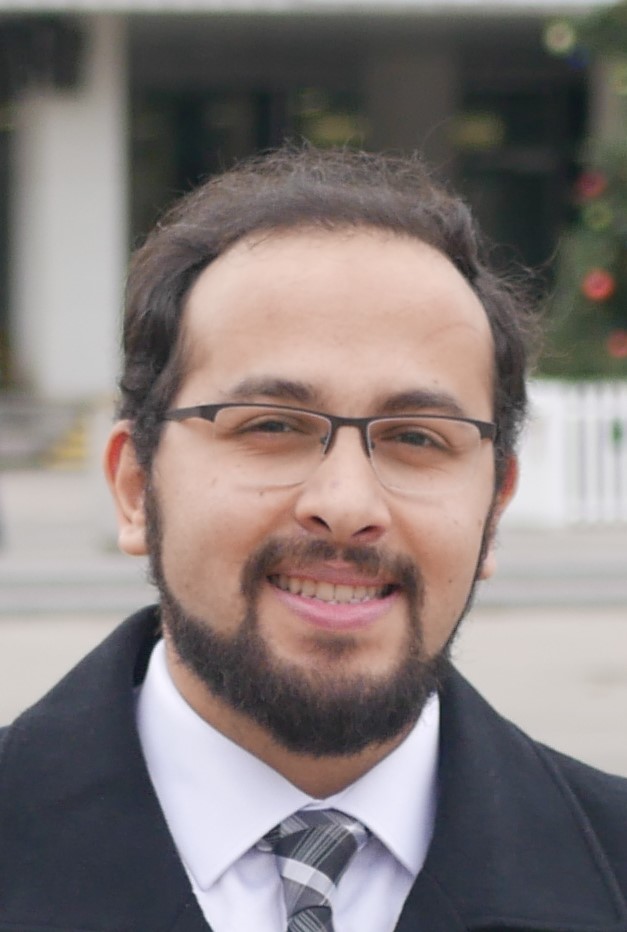}}]{Khaled Ahmed}
\:received his B.S. and MSc degree on wireless communications from Cairo university, Egypt, in 2010 and 2015. He did a PhD and a postdoc in McMaster university, Canada, in 2019 and 2021. His PhD degree was on optical communications while his postdoc was on machine learning for resource-limited devices. \par 
He is currently working as a Member Technical Staff II for MDA, Sainte-Anne-de Bellevue, QC, Canada. \par
His current research interests include RF and optical communications for terrestrial and satellite networks, optical beamforming networks, and applied machine learning and deep learning.  \par
\end{IEEEbiography}

\begin{IEEEbiography}[{\includegraphics[width=1in,height=1.25in,clip,keepaspectratio]{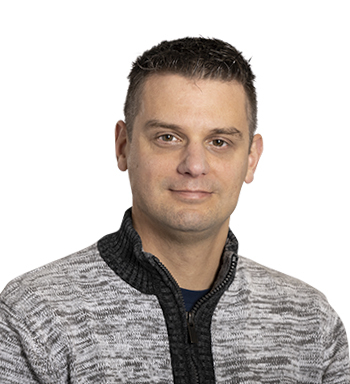}}]{Stéphane Martel}
\: received his Bachelor's degree in Electrical Engineering from McGill University. He is the Product Development Manager, Technology Strategy of Satellite Systems, at MDA Canada.
\end{IEEEbiography}

\end{document}